\documentclass[12pt]{article}
\usepackage{ctex}
\usepackage{amssymb,amsmath,color,mathdots}
\usepackage[colorlinks,linkcolor=blue]{hyperref}
\usepackage{geometry}
\usepackage{tabularx}
\usepackage{bm}
\usepackage{authblk}
\usepackage{graphicx}
\usepackage[square, comma, sort&compress, numbers]{natbib}

\title{\bf Complete weight enumerators for several classes of two-weight and three-weight linear codes}	
\author{\small Canze Zhu}
\author{\small Qunying Liao
	\thanks{Corresponding author.
		
		{~E-mail. qunyingliao@sicnu.edu.cn (Q. Liao), ~canzezhu@163.com (C. Zhu).}	
		
		{~Supported by National Natural Science Foundation of China (Grant No. 12071321).}}
}
\affil[]{\small College of Mathematical Science, Sichuan Normal University, Chengdu Sichuan, 610066}
\date{}
\newtheorem{theorem}{Theorem}[section]

\newtheorem{lemma}{Lemma}[section]
\newtheorem{example}{Example}[section]

\newtheorem{remark}{Remark}[section]

\catcode`@=11 \@addtoreset{equation}{section} \catcode`@=12
\begin{document}
	\maketitle
	{\bf Abstract.}
	{\small
		In this paper, for an odd prime $p$, by extending Li et al.'s construction \cite{CL2016}, several classes of two-weight and three-weight linear codes over the finite field $\mathbb{F}_p$ are constructed from a defining set, and then their complete weight enumerators are determined by using Weil sums. Furthermore, we show that some examples of these codes are optimal or almost optimal with respect to the Griesmer bound. Our results generalize the corresponding results in \cite{CL2016, GJ2019}.}\\
	
	\section{Introduction}
	Let $\mathbb{F}_{p^m}$ be the finite field with $p^m$ elements and $\mathbb{F}_{p^m}^*=\mathbb{F}_{p^m}\backslash \{0\}$, where $p$ is an odd prime and $m$ is a positive integer.   An $[n,k,d]$ linear code over $\mathbb{F}_p$ is a $k$-dimensional subspace of $\mathbb{F}_p^n$ with minimum distance $d$. In addition, the weight enumerator and complete weight enumerator are the important parameters for a linear code \cite{KT2007}, especially, few-weight linear codes have better applications \cite{AC1984,RC1986, JY2006,CD2005,CC2005}. Motivated by Ding et al's work \cite{KD2015}, a number of two-weight or three-weight linear codes have been constructed from defining sets \cite{KD2014, SY2015, ZH2015, ZH2016, ZH20161, ZH20162, GJ2019, CL2016, GL2018, CT2016, ZZ2015, CS2019}. 

	In 2015, Ding et al. gave a method to construct a class of two-weight or three-weight linear codes via the trace function from defining sets \cite{KD2015}. Let 
	$D=\{ d_1, d_2,\cdots, d_n \}\subseteq\mathbb{F}_{p^{m}}^*$ and $\mathrm{Tr}_m$ denote the trace function from $\mathbb{F}_{p^{m}}$ onto $\mathbb{F}_p$, a $p$-ary linear code is defined by
	\begin{align*}
	\mathcal{C}_D=\big\{\mathbf{c}(x)=\big(\mathrm{Tr}_m(xd_1),\mathrm{Tr}_m(xd_2),\cdots,\mathrm{Tr}_m(xd_n)\big)~\big{|}~ x\in\mathbb{F}_p^{m}\big\}.
	\end{align*} 
	Motivated by the above construction, Li et al. defined a linear code
	\begin{align}\label{C}
	\mathcal{C}_{\bar{D}}=\big\{\mathbb{c}(a,b)=\big(\mathrm{Tr}_m(ax+by)\big)_{(x,y)\in D}~\big{|}~ (a,b)\in\mathbb{F}_{p^m}\times\mathbb{F}_{p^m}\big\}
	\end{align}
	with $\bar{D}\subseteq \mathbb{F}_{p^m}^2$ \cite{CL2016}. Later, Jian et al. obtained several classes of two-weight and three-weight linear codes $\mathcal{C}_{\bar{D}}$ from $(\ref{C})$ by choosing the defining set 
	\begin{align}\label{D_1}
		\bar{D}=\big\{(x,y)\in\mathbb{F}_{p^m}^2 ~\big{|}~\mathrm{Tr}_m(x^2+y^{p^u+1})=0\big\},
	\end{align} where $u$ is a positive integer \cite{GJ2019}. 
	
	In this paper, 	we define a linear code 
	\begin{align}\label{CD}
	\mathcal{C}_{D_\lambda}=\big\{\big(\mathrm{Tr}_{m_1}(ax)+\mathrm{Tr}_{m_2}(by)\big)_{(x,y)\in D_\lambda}~\big{|}~(a,b)\in\mathrm{F}_{p^{m_1}}\times\mathrm{F}_{p^{m_2}}\big\}	
	\end{align}	with
	\begin{align}\label{D}
	D_\lambda
	=&\big\{(x,y)\in\mathbb{F}_{p^{m_1}}\times\mathbb{F}_{p^{m_2}}\backslash\{(0,0)\} ~\big{|}~\mathrm{Tr}_{m_1}(x^2)+\mathrm{Tr}_{m_2}(y^{p^u+1})=\lambda\big\},
	\end{align}
	where $\lambda\in\mathbb{F}_p$, $m_1$ and $m_2$ are positive integers.
	We determine the parameters and the complete weight enumerators of $\mathcal{C}_{D_\lambda}$ basing on Weil sums. In
	addition, for some examples, $\mathcal{C}_{D_\lambda}$ is optimal or almost optimal
	with respect to the Griesmer bound \cite{JH1960}. Obviously, 
	if $m_1=m_2$ and $\lambda=0$, then $D_\lambda=\bar{D}$ and $\mathcal{C}_{D_\lambda}=\mathcal{C}_{\bar{D}}$. Thus, we extend Li et al.'s construction \cite{CL2016}, and  generalize the corresponding  results in \cite{CL2016, GJ2019}.
	
	This paper is organized as follows. In section 2, some related basic notations and
	results of Weil sums are given. In section 3, the complete weight enumerators of several
	classes of two-weight and three-weight linear codes are presented. In section 4, the proofs of the main results are given. In
	section 5,  some examples are obtained by using Magma, which are accordant with the main results. In
	section 6, we conclude the whole paper.
	\section{Preliminaries}
   	Throughout the paper, we denote some notations as follows.\\

	$\bullet$ $\zeta_p=e^{\frac{2\pi\sqrt{-1}}{p}}$ is a primitive $p$-th root of the unity.\\
	
	$\bullet$  $u$, $m_1$ and $m_2$ are positive integers, and $s=\frac{m_2}{2}$, $v=\mathrm{gcd}(m_2, u)$.\\
	
	$\bullet$ $K=m_1+m_2$.\\
	
	$\bullet$ $L=(-1)^{\frac{(p-1)^2}{8}}$. \\
	
 	$\bullet$  For each $b\in \mathbb{F}_{p^m}$, $\chi_b(x)=\zeta_p^{\mathrm{Tr}_m(bx)}$ $(x\in \mathbb{F}_{p^m})$ is the addtive characters.\\
 	
 	$\bullet$ $\eta_m$ is the quadratic characters of $\mathbb{F}_{p^m}$, and it is extended by letting $\eta_m(0)=0$.\\
 	
 	$\bullet$ $G_m$ is the quadratic Gauss sums over $\mathbb{F}_{p^m}$, i.e., $G_m=\sum\limits_{c\in \mathbb{F}_{p^m}}\eta_m(c)\chi_1(c)$.
	\subsection{Group characters and Gauss sums}

	In this subsection, some properties for the additive characters, quadratic characters and quadratic Gauss sums are given.
	\begin{lemma}[\cite{KD2015,RL97}]\label{l21}
		$\quad G_m=(-1)^{m-1}L^mp^{\frac{m}{2}}$,
		and for $b\in\mathbb{F}_{p^m}$, 
		\begin{align*}
		\sum_{x\in \mathbb{F}_{p^m}}\zeta_p^{\mathrm{Tr}_m(bx)}=\begin{cases}
		p^m,\quad& ~b=0;\\
		0,\quad &\text{otherwise};
		\end{cases}
		\end{align*} 
	    for $x\in\mathbb{F}_p^*$, 
    	\begin{align*}
    	\eta_m(x)=\begin{cases}
    	1,\quad&  2~|~m;\\
    	\eta_1(x),\quad &otherwise.
    	\end{cases}
    	\end{align*}     	
    \end{lemma}

\subsection{Weil sums}
   It is well known that Weil sums are defined by $\sum\limits_{x\in\mathbb{F}_{p^{m_2}}}\chi(f(x))$ with $f(x)\in\mathbb{F}_{p^{m_2}}[x]$, and there are many results for the Weil sum \cite{CR,CR1}
\begin{align*}
S_{m_2,u}(a,b)=\sum_{x\in\mathbb{F}_{p^{m_2}}}\chi(ax^{p^u+1}+bx)~ (a\in\mathbb{F}_{p^{m_2}}^*,~b\in\mathbb{F}_{p^{m_2}}).
\end{align*}
\begin{lemma}\label{l23}
	If $\frac{m_2}{v}$ is odd, then
	\begin{align*}
	S_{m_2,u}(a,0)=G_{m_2}\eta_{m_2}(a).
	\end{align*}	
	\label{l24}
	If $\frac{m_2}{v}$ is even, then
	\begin{align*}
	S_{m_2,u}(a,0)=\begin{cases}
	p^s,\quad&\text{if}~\frac{s}{v}~\text{is even and}~a^{\frac{p^{m_2}-1}{p^v+1}}\neq(-1)^{\frac{s}{v}} ;\\
	-p^{s+v},\quad&\text{if}~\frac{s}{v}~\text{is even and}~a^{\frac{p^{m_2}-1}{p^v+1}}=(-1)^{\frac{s}{v}} ;\\
	-p^s,\quad&\text{if}~\frac{s}{v}~\text{is odd and}~a^{\frac{p^{m_2}-1}{p^v+1}}\neq(-1)^{\frac{s}{v}} ;\\
	p^{s+v},\quad&\text{if}~\frac{s}{v}~\text{is odd and}~a^{\frac{p^{m_2}-1}{p^v+1}}=(-1)^{\frac{s}{v}} .
	\end{cases}
	\end{align*}
\end{lemma}
\begin{lemma}\label{l25} Fixed $a\in\mathbb{F}_{p^{m_2}}$, then the equation \begin{align*}
	a^{p^u}X^{p^{2u}}+aX=0
	\end{align*}
	is solvable in $\mathbb{F}_{p^{m_2}}^{*}$ if and only if both $\frac{m_2}{v}$ is even and $a^{\frac{p^{m_2}-1}{p^v+1}}=(-1)^{\frac{s}{v}}$. Furthermore, there are exactly $p^{2v}-1$ non-zero solutions in this case.
\end{lemma}
\begin{remark}
	By Lemma $\ref{l25}$, it is easy to see that $f(X)=a^{p^u}X^{p^{2u}}+aX$ is a permutation polynomial over $\mathbb{F}_{p^{m_2}}^{*}$ if and only if $\frac{m_2}{v}$ is odd, or both $\frac{m_2}{v}$ is even and $a^{\frac{p^{m_2}-1}{p^v+1}}\neq(-1)^{\frac{s}{v}}$.
\end{remark}

\begin{lemma}\label{l26}
	Suppose that $f(X) =a^{p^u}X^{p^{2u}}+aX$ is a permutation polynomial over $\mathbb{F}_{p^{m_2}}$, then, $f(X)=-b^{p^u}$ has an unique solution in $\mathbb{F}_{p^{m_2}}$. Furthermore,
	\begin{align*}
	S_{m_2,u}(a,b)=	\begin{cases}
		G_{m_2}\eta_{m_2}(a)\zeta_p^{\mathrm{Tr_{m_2}}(-ax_0^{p^u+1})},  \qquad&\text{if}~\frac{m_2}{v}~\text{is odd};\\
		(-1)^{\frac{s}{v}}p^s\zeta_p^{\mathrm{Tr_{m_2}}(-ax_0^{p^u+1})}, \qquad&\text{if}~\frac{m_2}{v}~ \text{is even}.
		\end{cases}
	\end{align*}
\end{lemma}  
\begin{lemma}\label{l27}
	For the non-permutation polynomial $f(X) =a^{p^u}X^{p^{2u}}+aX$ over $\mathbb{F}_{p^{m_2}}$, suppose that the equation $f(X)=- b^{p^u}$ has a solution $x_0$ in $\mathbb{F}_{p^{m_2}}$, then,
	\begin{align*}
	S_{m_2,u}(a,b)=-(-1)^{\frac{s}{v}}p^{s+v}\zeta_p^{\mathrm{Tr_{m_2}}(-ax_0^{p^u+1})},
	\end{align*}	
	otherwise, 
	\begin{align*}
	S_{m_2,u}(a,b)=0.
	\end{align*}
\end{lemma}  

Taking $u = 0$ in Lemmas \ref{l23} and \ref{l26}, we can get
\begin{lemma}\label{l28}
	For $a\in\mathbb{F}_{p^{m_1}}^*$ and $b\in\mathbb{F}_{p^{m_1}}$, 
	\begin{align*}
	Q_{m_1}(a,b)
	=\sum_{x\in{\mathbb{F}_p^{m_1}}}\zeta_p^{\mathrm{Tr_{m_1}}(ax^2+bx)}=G_{m_1}\eta_{m_1}(a)\zeta_p^{\mathrm{Tr_{m_1}}(-\frac{b^2}{4a})}.
	\end{align*}
\end{lemma}    

   In order to prove our main results, we need the evaluation of $S_{m_2,u}(z_1,z_2b)$, where  $z_1,z_2\in\mathbb{F}_{p}^{*}$ and $b\in\mathbb{F}_{p^{m_2}}$. The following lemma is necessary.
\begin{lemma}[\cite{GJ2019}, Lemma 10]\label{l29}
	If $z\in\mathbb{F}_p^{*}$ and $\frac{m_2}{v}$ is even, then
	\begin{align*}
	z^{\frac{p^{m_2}-1}{p^v+1}}=1.
	\end{align*}
\end{lemma} 

By Lemmas \ref{l25} and \ref{l29}, the equation
\begin{align}\label{E1}
X^{p^{2u}}+X=-b^{p^u}
\end{align}is not always solvable in $\mathbb{F}_{p^{m_2}}$ when  $\frac{m_2}{v}\equiv 0~(\mathrm{mod}~4)$ and has a unique solution otherwise. Now, suppose that $(\ref{E1})$ has a solution $\gamma_b\in\mathbb{F}_{p^{m_2}}$, then, $\frac{z_2}{z_1}\gamma_b$ is a solution of the equation
\begin{align}\label{E2}
z_1^{p^{u}}X^{p^{2u}}+z_1X=-(z_2b)^{p^{u}}.
\end{align}
Thus, by Lemmas $\ref{l25}$-$\ref{l28}$, the evaluation of $S_{m_2,u}(z_1,z_2b)$ is given in the following 
\begin{lemma}\label{lS1}
	For $z_1,z_2\in\mathbb{F}_{p}^{*}$ and $b\in\mathbb{F}_{p^{m_2}}$,
	\begin{align*}
	S_{m_2,u}(z_1,z_2b)=\begin{cases}
	G_{m_2}\eta_{m_2}(z_1)\zeta_p^{-\frac{z_2^2}{z_1}\mathrm{Tr}_{m_2}(\gamma_b^{p^{u}+1})},\quad&\text{if}~\frac{m_2}{v} \text{  is odd};\\
	-p^s\zeta_p^{-\frac{z_2^2}{z_1}\mathrm{Tr}_{m_2}(\gamma_b^{p^{u}+1})},\quad&\text{if}~\frac{m_2}{v}\equiv 2~(\mathrm{mod}~4);\\
	-p^{s+v}\zeta_p^{-\frac{z_2^2}{z_1}\mathrm{Tr}_{m_2}(\gamma_b^{p^{u}+1})} ,\quad&\text{if}~\frac{m_2}{v}\equiv 0~(\mathrm{mod}~4)~\text{and}~(\ref{E1})~\text{is solvable};\\
	0,\quad&\text{if}~\frac{m_2}{v}\equiv 0~(\mathrm{mod}~4)~\text{and}~(\ref{E1})~\text{is not solvable}.\\
	\end{cases}
	\end{align*}
\end{lemma}

Especially, for $b=0$, $\gamma_0=0$ is a solution of $(\ref{E2})$. Thus, we have the following       
\begin{lemma}\label{lS2}
	For $z_1\in\mathbb{F}_p^{*}$,
	\begin{align*}
	S_{m_2,u}(z_1,0)=\begin{cases}
	G_{m_2}\eta_{m_2}(z_1),\quad&\text{if}~\frac{m_2}{v} \text{  is odd};\\
	-p^s,\quad&\text{if}~\frac{m_2}{v}\equiv 2~(\mathrm{mod}~4);\\
	-p^{s+v},\quad&\text{if}~\frac{m_2}{v}\equiv 0~(\mathrm{mod}~4).\\
	\end{cases}
	\end{align*}
\end{lemma}
\subsection{The Pless power moments}
The following lemma is necesssary to calculate the  weight enumerator of $\mathcal{C}_{D_\lambda}$.


	\begin{lemma}[\cite{WV}, p.259, The Pless power moments]\label{l12}
	For an $[n, k, d]$ code $\mathcal{C}$ over $\mathbb{F}_p$ with weight distribution $(1, A_1 , ..., A_n)$, suppose that
	the weight distribution of  its dual code is $(1, A_1^{\bot}, ..., A_n^{\bot})$, then the first two Pless power
	moments are
	\begin{align*}
	\sum_{j=0}^{n}A_j=p^{k}
	\end{align*}and
	\begin{align*}
	\sum_{j=0}^{n}jA_j=p^{k-1}(pn-n- A_1^{\bot}).
	\end{align*}	
	\end{lemma}

For $C_{D_\lambda}$ defined by $(\ref{CD})$, if $(0,0)\notin D_\lambda$, by the nondegenerate property of the trace function, one has $A_1^{\perp}=0$.	
\section{Main results}		
In this subsection, for $D_\lambda$ and $\mathcal{C}_{D_\lambda}$ given by $(\ref{D})$ and  $(\ref{CD})$, respectively, we present the complete weight enumerators of $\mathcal{C}_{D_\lambda}$ by classfying $\lambda=0$ or not, $m_1$ is odd or even, and $\frac{m_2}{v}~\mathrm{mod}~4$. 
Furthermore, for any given $c\in \mathbb{F}_{p}^{*}$,
\begin{align*}
\mathrm{Tr}_{m_1}\big((cx)^2\big)+\mathrm{Tr}_{m_2}\big((cy)^{p^u+1}\big)=c^2\big(\mathrm{Tr}_{m_1}(x^2)+\mathrm{Tr}_{m_2}(y^{p^u+1})\big),
\end{align*} 
hence, $D_0$ can be expressed as
\begin{align}\label{DD}
{D_0}=\cup_{c\in\mathbb{F}_p^{*}}\tilde{D}_0
\end{align} with $\tilde{D}_0\subsetneq{D_0}$. Thus, $\mathcal{C}_{\tilde{D}_0}$ defined by $(\ref{CD})$ is just the punctured version  of $\mathcal{C}_{D_0}$.

For $\lambda\in\mathbb{F}_{p}^{*}$, since
\begin{align*}
\mathrm{Tr}_{m_1}\big((-x)^2\big)+\mathrm{Tr}_{m_2}\big((-y)^{p^u+1}\big)=\mathrm{Tr}_{m_1}(x^2)+\mathrm{Tr}_{m_2}(y^{p^u+1}),
\end{align*} 
and then, 
\begin{align}\label{Dl}
{D_\lambda}=\cup_{c\in\mathbb{F}_p^{*}}\tilde{D}_\lambda\end{align} with $\tilde{D}_\lambda\subsetneq{D_\lambda}$. Thus, $\mathcal{C}_{\tilde{D}_\lambda}$ defined by $(\ref{CD})$ is just the punctured version  of $\mathcal{C}_{D_\lambda}$.

 The parameters of $\mathcal{C}_{D_\lambda}$ and $\mathcal{C}_{\tilde{D}_\lambda}$ are given in the following
theorems.
\begin{theorem}\label{t1}
If $\frac{m_2}{v}$ and $K$ are both odd, or $\frac{m_2}{v}\equiv 2~(\mathrm{mod}~4)$ and $m_1$ is odd, then $\mathcal{C}_{D_0}$ is a $\big{[}p^{K-1}-1, K\big{]}$ code with weight enumerator in Table $1$, and the complete weight enumerator is 
	\begin{align}\label{w1}
	\begin{aligned}
	\mathrm{W}(\mathcal{C}_{D_0})= &w_0^{p^{K-1}-1}+\big(p^{K-1}-1\big)w_0^{p^{K-2}-1}\prod_{i\in \mathbb{F}_p^*}w_i^{p^{K-2}}\\&+\frac{p-1}{2}p^{\frac{K-1}{2}}\big(p^{\frac{K-1}{2}}+1\big)w_0^{p^{K-2}+(p-1)p^{\frac{K-3}{2}}}\prod_{i\in \mathbb{F}_p^*}w_i^{p^{K-2}-p^{\frac{K-3}{2}}}
	\\&+\frac{p-1}{2}p^{\frac{K-1}{2}}\big(p^{\frac{K-1}{2}}-1\big)w_0^{p^{K-2}-(p-1)p^{\frac{K-3}{2}}}\prod_{i\in \mathbb{F}_p^*}w_i^{p^{K-2}+p^{\frac{K-3}{2}}}.
	\end{aligned}
	\end{align}	
	
	\begin{center} Table $1$~~~ The weight enumerator of  $\mathcal{C}_{D_0}$
		
		\begin{tabular}{|p{4cm}<{\centering}| p{4cm}<{\centering}|}
			\hline   weight $w$	                     &   frequency $A_w$                    \\ 
			\hline       $0$	                     &  $1$                                    \\ 
			\hline   $(p-1)p^{K-2}$   &  $(p^{K-1}-1)$        \\ 
			\hline   $(p-1)(p^{K-2}-p^{\frac{K-3}{2}})$   &  $\frac{p-1}{2}p^{\frac{K-1}{2}}(p^{\frac{K-1}{2}}+1)$        \\ 
			\hline   $(p-1)(p^{K-2}+p^{\frac{K-3}{2}})$	                     &      $\frac{p-1}{2}p^{\frac{K-1}{2}}(p^{\frac{K-1}{2}}-1)$  \\         
			\hline
		\end{tabular} 
	\end{center} Furthermore, $\mathcal{C}_{\tilde{D}_0}$ is a $\big{[}\frac{p^{K-1}-1}{p-1}, K\big{]}$ code with weight enumerator in Table $1^{\circ}$.
\begin{center} Table $1^{\circ}$~~~ The weight enumerator of  $\mathcal{C}_{\tilde{D}_0}$
	
	\begin{tabular}{|p{4cm}<{\centering}| p{4cm}<{\centering}|}
		\hline   weight $w$	                     &   frequency $A_w$                    \\ 
		\hline       $0$	                     &  $1$                                    \\ 
		\hline   $p^{K-2}$   &  $(p^{K-1}-1)$        \\ 
		\hline   $p^{K-2}-p^{\frac{K-3}{2}}$   &  $\frac{p-1}{2}p^{\frac{K-1}{2}}(p^{\frac{K-1}{2}}+1)$        \\ 
		\hline   $p^{K-2}+p^{\frac{K-3}{2}}$	                     &      $\frac{p-1}{2}p^{\frac{K-1}{2}}(p^{\frac{K-1}{2}}-1)$  \\         
		\hline
	\end{tabular} 
\end{center} 
\end{theorem}
\begin{theorem}\label{t2}
If $\frac{m_2}{v}$ is odd and $K$ is even, then $\mathcal{C}_{D_0}$ is a $\big{[}p^{K-1}+{L^{K}}(p-1)p^{\frac{K-2}{2}}-1, K\big{]}$ code with weight enumerator in Table $2$, and  the complete weight enumerator is 
	\begin{align}\label{w2}
	\begin{aligned}
	&\mathrm{W}(\mathcal{C}_{D_0})\\= &w_0^{p^{K-1}+{L^{K}}(p-1)p^{\frac{K-2}{2}}-1}+(p-1)p^{\frac{K-2}{2}}\big(p^{\frac{K}{2}}-{L^{K}}\big)w_0^{p^{K-2}-1}\prod_{i\in \mathbb{F}_p^*}w_i^{p^{\frac{K-2}{2}}\big(p^{\frac{K-2}{2}}+{L^{K}}\big)}\\&+\big(p^{K-1}+{L^{K}}(p-1)p^{\frac{K-2}{2}}-1\big)w_0^{p^{K-2}+{L^{K}}(p-1)p^{\frac{K-2}{2}}-1}\prod_{i\in \mathbb{F}_p^*}w_i^{p^{K-2}}.
	\end{aligned}
	\end{align}	
	
	\begin{center} Table $2$~~~ The weight enumerator of  $\mathcal{C}_{D_0}$
		
		\begin{tabular}{|p{4.5cm}<{\centering}| p{4.5cm}<{\centering}|}
			\hline   weight $w$	                     &   frequency $A_w$                    \\ 
			\hline       $0$	                     &  $1$                                    \\
			\hline   $(p-1)p^{\frac{K-2}{2}}(p^{\frac{K-2}{2}}+{L^{K}})$   &  $(p-1)p^{\frac{K-2}{2}}(p^{\frac{K}{2}}-{L^{K}})$        \\ 
			\hline   $(p-1)p^{K-2}$	                     &      $p^{K-1}+{L^{K}}(p-1)p^{\frac{K-2}{2}}-1$  \\         
			\hline
		\end{tabular} 
	\end{center} 
 Furthermore, $\mathcal{C}_{\tilde{D}_0}$ is a $\big{[}\frac{p^{K-1}+-1}{p-1}+{L^{K}}p^{\frac{K-2}{2}}, K\big{]}$ code with weight enumerator in Table $2^{\circ}$.
	
	\begin{center} Table $2^{\circ}$~~~ The weight enumerator of  $\mathcal{C}_{\tilde{D}_0}$
		
		\begin{tabular}{|p{4.5cm}<{\centering}| p{4.5cm}<{\centering}|}
			\hline   weight $w$	                     &   frequency $A_w$                    \\ 
			\hline       $0$	                     &  $1$                                    \\
			\hline   $p^{\frac{K-2}{2}}(p^{\frac{K-2}{2}}+{L^{K}})$   &  $(p-1)p^{\frac{K-2}{2}}(p^{\frac{K}{2}}-{L^{K}})$        \\ 
			\hline   $p^{K-2}$	                     &      $p^{K-1}+{L^{K}}(p-1)p^{\frac{K-2}{2}}-1$  \\         
			\hline
		\end{tabular} 
	\end{center} 
\end{theorem}
\begin{theorem}\label{t3}
If $\frac{m_2}{v}\equiv 2~(\mathrm{mod}~4)$ and $m_1$ is even, then $\mathcal{C}_{D_0}$ is a $\big{[}p^{K-1}+{L^{m_1}}(p-1)p^{\frac{K-2}{2}}-1, K\big{]}$ code with weight enumerator in Table $3$, and  the complete weight enumerator is 
	\begin{align}\label{w3}
	\begin{aligned}
	&\mathrm{W}(\mathcal{C}_{D_0})\\= &w_0^{p^{K-1}+{L^{m_1}}(p-1)p^{\frac{K-2}{2}}-1}+(p-1)p^{\frac{K-2}{2}}\big(p^{\frac{K}{2}}-{L^{m_1}}\big)w_0^{p^{K-2}-1}\prod_{i\in \mathbb{F}_p^*}w_i^{p^{\frac{K-2}{2}}\big(p^{\frac{K-2}{2}}+{L^{m_1}}\big)}\\&+\big(p^{K-1}+{L^{m_1}}(p-1)p^{\frac{K-2}{2}}-1\big)w_0^{p^{K-2}+{L^{m_1}}(p-1)p^{\frac{K-2}{2}}-1}\prod_{i\in \mathbb{F}_p^*}w_i^{p^{K-2}}.
	\end{aligned}
	\end{align}	
	
	\begin{center} Table $3$~~~ The weight enumerator of  $\mathcal{C}_{D_0}$
		
		\begin{tabular}{|p{4.5cm}<{\centering}| p{4.5cm}<{\centering}|}
			\hline   weight $w$	                     &   frequency $A_w$                    \\ 
			\hline       $0$	                     &  $1$                                    \\
			\hline   $(p-1)p^{\frac{K-2}{2}}(p^{\frac{K-2}{2}}+{L^{m_1}})$   &  $(p-1)p^{\frac{K-2}{2}}(p^{\frac{K}{2}}-{L^{m_1}})$        \\ 
			\hline   $(p-1)p^{K-2}$	                     &      $p^{K-1}+{L^{m_1}}(p-1)p^{\frac{K-2}{2}}-1$  \\         
			\hline
		\end{tabular} 
	\end{center} 
Furthermore, $\mathcal{C}_{\tilde{D}_0}$ is a $\big{[}\frac{p^{K-1}-1}{p-1}+{L^{m_1}}p^{\frac{K-2}{2}}, K\big{]}$ code with weight enumerator in Table $3^{\circ}$.
	
	\begin{center} Table $3^{\circ}$~~~ The weight enumerator of  $\mathcal{C}_{\tilde{D}_0}$
		
		\begin{tabular}{|p{4.5cm}<{\centering}| p{4.5cm}<{\centering}|}
			\hline   weight $w$	                     &   frequency $A_w$                    \\ 
			\hline       $0$	                     &  $1$                                    \\
			\hline   $p^{\frac{K-2}{2}}(p^{\frac{K-2}{2}}+{L^{m_1}})$   &  $(p-1)p^{\frac{K-2}{2}}(p^{\frac{K}{2}}-{L^{m_1}})$        \\ 
			\hline   $p^{K-2}$	                     &      $p^{K-1}+{L^{m_1}}(p-1)p^{\frac{K-2}{2}}-1$  \\         
			\hline
		\end{tabular} 
	\end{center} 
\end{theorem}
\begin{theorem}\label{t4}
If $\frac{m_2}{v}\equiv 0~(\mathrm{mod}~4)$ and $m_1$ is even, then $\mathcal{C}_{D_0}$ is a $\big{[}p^{K-1}+{L^{m_1}}(p-1)p^{\frac{K-2}{2}+v}-1,K\big{]}$ code with weight enumerator in Table $4$, and  the complete weight enumerator is 
	\begin{align}\label{w4}
	\begin{aligned}
	&\mathrm{W}(\mathcal{C}_{D_0})\\= &w_0^{p^{K-1}+{L^{m_1}}(p-1)p^{\frac{K-2}{2}+v}-1}+(p-1)p^{\frac{K-2}{2}-v}\big(p^{\frac{K}{2}-v}-{L^{m_1}}\big)w_0^{p^{K-2}-1}\prod_{i\in \mathbb{F}_p^*}w_i^{p^{K-2}+{L^{m_1}}p^{\frac{K-2}{2}+v}}\\&+\big(p^{K-2v-1}+
	{L^{m_1}}(p-1)p^{\frac{K-2}{2}-v}-1\big)w_0^{p^{K-2}+{L^{m_1}}(p-1)p^{\frac{K}{2}+v}-1}\prod_{i\in \mathbb{F}_p^*}w_i^{p^{K-2}}\\
	&+p^{K}\big(1-p^{-2v}\big)w_0^{p^{K-2}+{L^{m_1}}(p-1)p^{\frac{K-4}{2}+v}-1}\prod_{i\in \mathbb{F}_p^*}w_i^{p^{K-2}+{L^{m_1}}(p-1)p^{\frac{K-4}{2}+v}}.
	\end{aligned}
	\end{align}	
	
	\begin{center} Table $4$~~~ The weight enumerator of  $\mathcal{C}_{D_0}$
		
		\begin{tabular}{|p{6cm}<{\centering}| p{6cm}<{\centering}|}
			\hline   weight $w$	                     &   frequency $A_w$                    \\ 
			\hline       $0$	                     &  $1$                                    \\
			\hline   $(p-1)\big(p^{K-2}+{L^{m_1}}p^{\frac{K-2}{2}+v}\big)$   &  $(p-1)p^{\frac{K-2}{2}-v}\big(p^{\frac{K}{2}-v}-{L^{m_1}}\big)$        \\ 
			\hline   $(p-1)p^{K-2}$	                     &      $p^{K-2v-1}+
			{L^{m_1}}(p-1)p^{\frac{K-2}{2}-v}-1$  \\         
			\hline   $(p-1)\big(p^{K-2}+{L^{m_1}}(p-1)p^{\frac{K-4}{2}+v}\big)$   &  $p^{K}\big(1-p^{-2v}\big)$        \\ 
			\hline
		\end{tabular} 
	\end{center} 
 Furthermore, $\mathcal{C}_{\tilde{D}_0}$ is a $\big{[}\frac{p^{K-1}-1}{p-1}+{L^{m_1}}p^{\frac{K-2}{2}+v},K\big{]}$ code with weight enumerator in Table $4^{\circ}$.

\begin{center} Table $4^{\circ}$~~~ The weight enumerator of  $\mathcal{C}_{\tilde{D}_0}$
	
	\begin{tabular}{|p{5cm}<{\centering}| p{6cm}<{\centering}|}
		\hline   weight $w$	                     &   frequency $A_w$                    \\ 
		\hline       $0$	                     &  $1$                                    \\
		\hline   $p^{K-2}+{L^{m_1}}p^{\frac{K-2}{2}+v}$   &  $(p-1)p^{\frac{K-2}{2}-v}\big(p^{\frac{K}{2}-v}-{L^{m_1}}\big)$        \\ 
		\hline   $p^{K-2}$	                     &      $p^{K-2v-1}+
		{L^{m_1}}(p-1)p^{\frac{K-2}{2}-v}-1$  \\         
		\hline   $p^{K-2}+{L^{m_1}}(p-1)p^{\frac{K-4}{2}+v}$   &  $p^{K}\big(1-p^{-2v}\big)$        \\ 
		\hline
	\end{tabular} 
\end{center} 
\end{theorem}
\begin{theorem}\label{t5}
If  $\frac{m_2}{v}\equiv 0~(\mathrm{mod}~4)$ and $m_1$ is odd, then $\mathcal{C}_{D_0}$ is a $\big{[}p^{K-1}-1, K\big{]}$ code with weight enumerator in Table $5$, and  the complete weight enumerator is 
	\begin{align}\label{w5}
	\begin{aligned}
	&\mathrm{W}(\mathcal{C}_{D_0})\\= &w_0^{p^{K-1}-1}+\big(p^{K}-(p-1)p^{K-2v-1}-1\big)w_0^{p^{K-2}-1}\prod_{i\in \mathbb{F}_p^*}w_i^{p^{K-2}}\\&+\frac{p-1}{2}\big(p^{K-2v-1}-{L^{m_1+1}}p^{\frac{K-1}{2}-v}\big)w_0^{p^{K-2}-(p-1)p^{\frac{K-3}{2}+v}-1}\prod_{i\in \mathbb{F}_p^*}w_i^{p^{K-2}+p^{\frac{K-3}{2}+v}}
	\\&+\frac{p-1}{2}\big(p^{K-2v-1}+{L^{m_1+1}}p^{\frac{K-1}{2}-v}\big)w_0^{p^{K-2}+(p-1)p^{\frac{K-3}{2}+v}-1}\prod_{i\in \mathbb{F}_p^*}w_i^{p^{K-2}-p^{\frac{K-3}{2}+v}}.
	\end{aligned}
	\end{align}	
	
	\begin{center} Table $5$~~~ The weight enumerator of  $\mathcal{C}_{D_0}$
		
		\begin{tabular}{|p{5cm}<{\centering}| p{7cm}<{\centering}|}
			\hline   weight $w$	                     &   frequency $A_w$                    \\ 
			\hline       $0$	                     &  $1$                                    \\ 
			\hline   $(p-1)p^{K-2}$   & $p^{K}-(p-1)p^{K-2v-1}-1$       \\ 
			\hline   $(p-1)(p^{K-2}+p^{\frac{K-3}{2}+v})$	                     &    $\frac{p-1}{2}\big(p^{K-2v-1}-\eta_{1}(-1){L^{m_1+1}}p^{\frac{K-1}{2}-v}\big)$ \\
			\hline   $(p-1)(p^{K-2}-p^{\frac{K-3}{2}+v})$   &  $\frac{p-1}{2}\big(p^{K-2v-1}+\eta_{1}(-1){L^{m_1+1}}p^{\frac{K-1}{2}-v}\big)$        \\        
			\hline
		\end{tabular} 
	\end{center}  Furthermore, $\mathcal{C}_{\tilde{D}_0}$ is a $\big{[}\frac{p^{K-1}-1}{p-1}, K\big{]}$ code with weight enumerator in Table $5^{\circ}$.
	
	\begin{center} Table $5^{\circ}$~~~ The weight enumerator of  $\mathcal{C}_{\tilde{D}_0}$
		
		\begin{tabular}{|p{4cm}<{\centering}| p{7cm}<{\centering}|}
			\hline   weight $w$	                     &   frequency $A_w$                    \\ 
			\hline       $0$	                     &  $1$                                    \\ 
			\hline   $p^{K-2}$   & $p^{K}-(p-1)p^{K-2v-1}-1$       \\ 
			\hline   $p^{K-2}+p^{\frac{K-3}{2}+v}$	                     &    $\frac{p-1}{2}\big(p^{K-2v-1}-\eta_{1}(-1){L^{m_1+1}}p^{\frac{K-1}{2}-v}\big)$ \\
			\hline   $p^{K-2}-p^{\frac{K-3}{2}+v}$   &  $\frac{p-1}{2}\big(p^{K-2v-1}+\eta_{1}(-1){L^{m_1+1}}p^{\frac{K-1}{2}-v}\big)$        \\        
			\hline
		\end{tabular} 
	\end{center} 
\end{theorem}
\newpage
\begin{theorem}\label{t6}
	For $\lambda\in \mathbb{F}_p^{*}$, if $\frac{m_2}{v}$ and $K$ are both odd, then $\mathcal{C}_{D_\lambda}$ is a ${[}p^{K-1}-\eta_{1}(-\lambda){{L^{K+1}}}p^{\frac{K-1}{2}},\\ K{]}$ code with weight enumerator in Table $6$, and  the complete weight enumerator is 
	{\footnotesize\begin{align}\label{w6}
		\begin{aligned}
		&\mathrm{W}(\mathcal{C}_{D_\lambda})\\
		= &w_0^{p^{K-1}-\eta_{1}(-\lambda){{L^{K+1}}}p^{\frac{K-1}{2}}}+\big(p^{K-1}-1\big)w_0^{p^{K-2}-\eta_{1}(-\lambda){{L^{K+1}}}p^{\frac{K-1}{2}}}\prod_{i\in \mathbb{F}_p^*}w_i^{p^{K-2}}\\&+\frac{p-1}{2}p^{\frac{K-1}{2}}\big(p^{\frac{K-1}{2}}+\eta_{1}(-\lambda){{L^{K+1}}}\big)\prod_{i\in \mathbb{F}_p}w_i^{p^{K-2}-\eta_{1}(-\lambda){{L^{K+1}}}p^{\frac{K-3}{2}}}	\\&+p^{\frac{K-1}{2}}(p^{\frac{K-1}{2}}-\eta_{1}(-\lambda){{L^{K+1}}})\sum_{\substack{j\in\mathrm{F}_p^{*}\\ \eta_1(j)=\eta_1(\lambda)}}
		w_{i=\pm\sqrt{4\lambda j}}^{p^{K-2}-\eta_{1}(-\lambda){{L^{K+1}}}(p-1)p^{\frac{K-3}{2}}}\prod_{\substack{i\in \mathbb{F}_p\\i\neq\pm\sqrt{4\lambda j}}}w_i^{p^{K-2}+\eta_{1}(-\lambda){{L^{K+1}}}p^{\frac{K-3}{2}}}. 
		\end{aligned}
		\end{align}	}
	\begin{center} Table $6$~~~ The weight enumerator of  $\mathcal{C}_{D_\lambda}$
		
		\begin{tabular}{|p{6.5cm}<{\centering}| p{6cm}<{\centering}|}
			\hline   weight $w$	                     &   frequency $A_w$                    \\ 
			\hline       $0$	                     &  $1$                                    \\ 
			\hline   $(p-1)p^{K-2}$   &  $(p^{K-1}-1)$        \\ 
			\hline   $(p-1)\big(p^{K-2}-\eta_{1}(-\lambda){{L^{K+1}}}p^{\frac{K-3}{2}}\big)$   &  $\frac{p-1}{2}p^{\frac{K-1}{2}}(p^{\frac{K-1}{2}}+\eta_{1}(-\lambda){{L^{K+1}}})$      \\ 
			\hline   $(p-1)p^{K-2}-\eta_{1}(-\lambda)(p+1){{L^{K+1}}}p^{\frac{K-3}{2}}$   	                     &      $\frac{p-1}{2}p^{\frac{K-1}{2}}(p^{\frac{K-1}{2}}-\eta_{1}(-\lambda){{L^{K+1}}})$  \\         
			\hline
		\end{tabular} 
	\end{center} Furthermore, $\mathcal{C}_{\tilde{D}_\lambda}$ is a $\big{[}\frac{1}{2}\big(p^{K-1}-\eta_{1}(-\lambda){{L^{K+1}}}p^{\frac{K-1}{2}}\big), K\big{]}$ code with weight enumerator in Table $6^{\circ}$.

\begin{center} Table $6^{\circ}$~~~ The weight enumerator of  $\mathcal{C}_{D_0}$
	
	\begin{tabular}{|p{6cm}<{\centering}| p{6cm}<{\centering}|}
		\hline   weight $w$	                     &   frequency $A_w$                    \\ 
		\hline       $0$	                     &  $1$                                    \\ 
		\hline   $\frac{p-1}{2}p^{K-2}$   &  $(p^{K-1}-1)$        \\ 
		\hline   $\frac{p-1}{2}\big(p^{K-2}-\eta_{1}(-\lambda){{L^{K+1}}}p^{\frac{K-3}{2}}\big)$   &  $\frac{p-1}{2}p^{\frac{K-1}{2}}(p^{\frac{K-1}{2}}+\eta_{1}(-\lambda){{L^{K+1}}})$      \\ 
		\hline   $\frac{p-1}{2}p^{K-2}-\frac{p+1}{2}\eta_{1}(-\lambda){{L^{K+1}}}p^{\frac{K-3}{2}}$   	                     &      $\frac{p-1}{2}p^{\frac{K-1}{2}}(p^{\frac{K-1}{2}}-\eta_{1}(-\lambda){{L^{K+1}}})$  \\         
		\hline
	\end{tabular} 
\end{center} 
\end{theorem}
\begin{theorem}\label{t7}
	For $\lambda\in \mathbb{F}_p^{*}$, if $\frac{m_2}{v}$ is odd and $K$ is even, then $\mathcal{C}_{D_\lambda}$ is a $\big{[}p^{K-1}-{L^{K}}p^{\frac{K-2}{2}}, K\big{]}$ code with weight enumerator in Table $7$, and  the complete weight enumerator is 
	\begin{align}\label{w7}
	\begin{aligned}
	\mathrm{W}(\mathcal{C}_{D_\lambda})= &w_0^{p^{K-1}-{L^{K}}p^{\frac{K-2}{2}}}+\big(p^{K-1}+(p-1)L^{K}p^{\frac{K-2}{2}}-1\big)w_0^{p^{K-2}-L^{K}p^{\frac{K-2}{2}}}
	\prod_{i\in \mathbb{F}_p^*}w_i^{p^{\frac{K-2}{2}}}\\&+\big(p^{K-1}-L^{K}p^{\frac{K-2}{2}}\big)
	\sum_{j \in \mathbb{F}_p^*}\prod_{i\in \mathbb{F}_p}w_{i}^{p^{K-2}-\eta_{1}(i^2-4\lambda j)L^{K}p^{\frac{K-2}{2}}}.
	\end{aligned}
	\end{align}		
	\begin{center} Table $7$~~~ The weight enumerator of  $\mathcal{C}_{D_\lambda}$
		\begin{tabular}{|p{4.5cm}<{\centering}| p{5cm}<{\centering}|}
			\hline   weight $w$	                     &   frequency $A_w$                    \\ 
			\hline       $0$	                     &  $1$                                    \\
			\hline   $(p-1)p^{K-2}$	                     &      $\frac{p+1}{2}p^{K-1}+\frac{p-1}{2}{L^{K}}p^{\frac{K-2}{2}}-1$  \\   		
			\hline   $(p-1)p^{K-2}-2{L^{K}}p^{\frac{K-2}{2}}$   &  $\frac{p-1}{2}p^{\frac{K-2}{2}}\big(p^{\frac{K}{2}}-{L^{K}}\big)$        \\       
			\hline
		\end{tabular}
	\end{center}\newpage \noindent Furthermore, $\mathcal{C}_{\tilde{D}_\lambda}$ is a $\big{[}\frac{1}{2}\big(p^{K-1}-{L^{K}}p^{\frac{K-2}{2}}\big), K\big{]}$ code with weight enumerator in Table $7^{\circ}$.
	\begin{center} Table $7^{\circ}$~~~ The weight enumerator of  $\mathcal{C}_{\tilde{D}_\lambda}$
		\begin{tabular}{|p{4cm}<{\centering}| p{5cm}<{\centering}|}
			\hline   weight $w$	                     &   frequency $A_w$                    \\ 
			\hline       $0$	                     &  $1$                                    \\
			\hline   $\frac{p-1}{2}p^{K-2}$	                     &      $\frac{p+1}{2}p^{K-1}+\frac{p-1}{2}{L^{K}}p^{\frac{K-2}{2}}-1$  \\   		
			\hline   $\frac{p-1}{2}p^{K-2}-{L^{K}}$   &  $\frac{p-1}{2}p^{\frac{K-2}{2}}\big(p^{\frac{K}{2}}-{L^{K}}\big)$        \\       
			\hline
		\end{tabular}
	\end{center} 
\end{theorem}
\begin{theorem}\label{t8}
	For $\lambda\in \mathbb{F}_p^{*}$, if $\frac{m_2}{v}\equiv 2~(\mathrm{mod}~ 4)$ and $K$ is odd, then $\mathcal{C}_{D_\lambda}$ is a $\big{[}p^{K-1}-\eta_{1}(-\lambda){{L^{m_1+1}}}p^{\frac{K-1}{2}}, K\big{]}$ code with weight enumerator in Table $8$, and  the complete weight enumerator is 
	{\footnotesize \begin{align}\label{w8}
		\begin{aligned}
		&\mathrm{W}(\mathcal{C}_{D_\lambda})\\
		= &w_0^{p^{K-1}-\eta_{1}(-\lambda){{L^{m_1+1}}}p^{\frac{K-1}{2}}}+\big(p^{K-1}-1\big)w_0^{p^{K-2}-\eta_{1}(-\lambda){{L^{m_1+1}}}p^{\frac{K-1}{2}}}\prod_{i\in \mathbb{F}_p^*}w_i^{p^{K-2}}\\&+\frac{p-1}{2}p^{\frac{K-1}{2}}\big(p^{\frac{K-1}{2}}-{{L^{m_1+1}}}(p-1)\big)\prod_{i\in \mathbb{F}_p}w_i^{p^{K-2}-\eta_{1}(-\lambda){{L^{m_1+1}}}p^{\frac{K-3}{2}}}	\\&+p^{\frac{K-1}{2}}(p^{\frac{K-1}{2}}+{{L^{m_1+1}}}(p-1)) \sum_{\substack{j\in\mathrm{F}_p^{*}\\ \eta_1(j)=\eta_1(\lambda)}}
		w_{i=\pm\sqrt{4\lambda j}}^{p^{K-2}-\eta_{1}(-\lambda){{L^{m_1+1}}}(p-1)p^{\frac{K-3}{2}}}\prod_{\substack{i\in \mathbb{F}_p^*\\i\neq\pm\sqrt{4\lambda j}}}w_i^{p^{K-2}+\eta_{1}(-\lambda){{L^{m_1+1}}}p^{\frac{K-3}{2}}}. 
		\end{aligned}
		\end{align}	}
	
	\begin{center} Table $8$~~~ The weight enumerator of  $\mathcal{C}_{D_\lambda}$
		
		\begin{tabular}{|p{7cm}<{\centering}| p{6cm}<{\centering}|}
			\hline   weight $w$	                     &   frequency $A_w$                    \\ 
			\hline       $0$	                     &  $1$                                    \\ 
			\hline   $(p-1)p^{K-2}$   &  $(p^{K-1}-1)$        \\ 
			\hline   $(p-1)(p^{K-2}-\eta_{1}(-\lambda){{L^{m_1+1}}}p^{\frac{K-3}{2}})$   &  $\frac{p-1}{2}p^{\frac{K-1}{2}}(p^{\frac{K-1}{2}}-{{L^{m_1+1}}}(p-1))$      \\ 
			\hline   $(p-1)p^{K-2}-\eta_{1}(-\lambda)(p+1){{L^{m_1+1}}}p^{\frac{K-3}{2}}$   	                     &      $\frac{p-1}{2}p^{\frac{K-1}{2}}(p^{\frac{K-1}{2}}+{{L^{m_1+1}}}(p-1))$  \\         
			\hline
		\end{tabular} 
	\end{center}  Furthermore, $\mathcal{C}_{\tilde{D}_\lambda}$ is a $\big{[}\frac{1}{2}\big(p^{K-1}-\eta_{1}(-\lambda){{L^{m_1+1}}}p^{\frac{K-1}{2}}\big), K\big{]}$ code with weight enumerator in Table $8^{\circ}$.

\begin{center} Table $8^{\circ}$~~~ The weight enumerator of  $\mathcal{C}_{\tilde{D}_\lambda}$
	
	\begin{tabular}{|p{6cm}<{\centering}| p{6cm}<{\centering}|}
		\hline   weight $w$	                     &   frequency $A_w$                    \\ 
		\hline       $0$	                     &  $1$                                    \\ 
		\hline   $\frac{p-1}{2}p^{K-2}$   &  $(p^{K-1}-1)$        \\ 
		\hline   $\frac{p-1}{2}\big(p^{K-2}-\eta_{1}(-\lambda){{L^{m_1+1}}}p^{\frac{K-3}{2}}\big)$   &  $\frac{p-1}{2}p^{\frac{K-1}{2}}(p^{\frac{K-1}{2}}-{{L^{m_1+1}}}(p-1))$      \\ 
		\hline   $\frac{p-1}{2}p^{K-2}+\eta_{1}(-\lambda)\frac{p+1}{2}{{L^{m_1+1}}}p^{\frac{K-3}{2}}$   	                     &      $\frac{p-1}{2}p^{\frac{K-1}{2}}(p^{\frac{K-1}{2}}+{{L^{m_1+1}}}(p-1))$  \\         
		\hline
	\end{tabular} 
\end{center} 
\end{theorem}
\begin{theorem}\label{t9}
	For $\lambda\in \mathbb{F}_p^{*}$, if $\frac{m_2}{v}\equiv 2~(\mathrm{mod}~4)$ and $K$ is even, then $\mathcal{C}_{D_\lambda}$ is a $\big{[}p^{K-1}-{L^{m_1}}p^{\frac{K-2}{2}}, K\big{]}$ code with weight enumerator in Table $9$, and  the complete weight enumerator is 
	\begin{align}\label{w9}
	\begin{aligned}
	\mathrm{W}(\mathcal{C}_{D_\lambda})= &w_0^{p^{K-1}-{L^{m_1}}p^{\frac{K-2}{2}}}+\big(p^{K-1}+(p-1)L^{m_1}p^{\frac{K-2}{2}}-1\big)w_0^{p^{K-2}-L^{m_1}p^{\frac{K-2}{2}}}
	\prod_{i\in \mathbb{F}_p^*}w_i^{p^{\frac{K-2}{2}}}\\&+\big(p^{K-1}-L^{m_1}p^{\frac{K-2}{2}}\big)
	\sum_{j \in \mathbb{F}_p^*}\prod_{i\in \mathbb{F}_p}w_{i}^{p^{K-2}-\eta_{1}(i^2-4\lambda j)L^{m_1}p^{\frac{K-2}{2}}}.
	\end{aligned}
	\end{align}		
	\begin{center} Table $9$~~~ The weight enumerator of  $\mathcal{C}_{D_\lambda}$
		\begin{tabular}{|p{5cm}<{\centering}| p{5cm}<{\centering}|}
			\hline   weight $w$	                     &   frequency $A_w$                    \\ 
			\hline       $0$	                     &  $1$                                    \\
			\hline   $(p-1)p^{K-2}$	                     &      $\frac{p+1}{2}p^{K-1}+\frac{p-1}{2}{L^{m_1}}p^{\frac{K-2}{2}}-1$  \\   		
			\hline   $(p-1)p^{K-2}-2{L^{m_1}}p^{\frac{K-2}{2}}$   &  $\frac{p-1}{2}p^{\frac{K-2}{2}}\big(p^{\frac{K}{2}}-{L^{m_1}}\big)$        \\       
			\hline
		\end{tabular}
	\end{center}  Furthermore, $\mathcal{C}_{\tilde{D}_\lambda}$ is a $\big{[}\frac{1}{2}\big(p^{K-1}-{L^{m_1}}p^{\frac{K-2}{2}}\big), K\big{]}$ code with weight enumerator in Table $9^{\circ}$.
	\begin{center} Table $9^{\circ}$~~~ The weight enumerator of  $\mathcal{C}_{\tilde{D}_\lambda}$
		\begin{tabular}{|p{4cm}<{\centering}| p{5cm}<{\centering}|}
			\hline   weight $w$	                     &   frequency $A_w$                    \\ 
			\hline       $0$	                     &  $1$                                    \\
			\hline   $\frac{p-1}{2}p^{K-2}$	                     &      $\frac{p+1}{2}p^{K-1}+\frac{p-1}{2}{L^{m_1}}p^{\frac{K-2}{2}}-1$  \\   		
			\hline   $\frac{p-1}{2}p^{K-2}-{L^{m_1}}p^{\frac{K-2}{2}}$   &  $\frac{p-1}{2}p^{\frac{K-2}{2}}\big(p^{\frac{K}{2}}-{L^{m_1}}\big)$        \\       
			\hline
		\end{tabular}
	\end{center} 
\end{theorem}
\begin{theorem}\label{t10}
	For $\lambda\in \mathbb{F}_p^{*}$, if $\frac{m_2}{v}\equiv 0~(\mathrm{mod}~4)$ and $m_1$ is even, then $\mathcal{C}_{D_\lambda}$ is a $\big{[}p^{K-1}-{L^{m_1}}p^{\frac{K-2}{2}+v},K\big{]}$ code with weight enumerator in Table $10$, and  the complete weight enumerator is 
	\begin{align}\label{w10}
	\begin{aligned}
	\mathrm{W}(\mathcal{C}_{D_\lambda})= &w_0^{p^{K-1}-{L^{m_1}}p^{\frac{K-2}{2}+v}}+\big(p^{K}-p^{K-2v}\big)
	\prod_{i\in \mathbb{F}_p}w_i^{p^{K-2}-{L^{m_1}}p^{\frac{K-4}{2}+v}}\\&+\big(p^{K-2v-1}+L^{m_1}(p-1)p^{\frac{K-2}{2}-v}-1\big)w_0^{p^{K-2}-{L^{m_1}}p^{\frac{K-2}{2}+v}}\prod_{i\in \mathbb{F}_p^*}w_i^{p^{K-2}}\\&\big(p^{K-2v-1}-{L^{m_1}}p^{\frac{K-2}{2}-v}\big)
	\sum_{j \in \mathbb{F}_p^*}\prod_{i\in \mathbb{F}_p}w_{i}^{p^{K-2}-\eta_{1}(i^2-4\lambda j)L^{K}p^{\frac{K-2}{2}}}.
	\end{aligned}
	\end{align}	
	
	\begin{center} Table $10$~~~ The weight enumerator of  $\mathcal{C}_{D_\lambda}$
		
		\begin{tabular}{|p{5.5cm}<{\centering}| p{6cm}<{\centering}|}
			\hline   weight $w$	                     &   frequency $A_w$                    \\ 
			\hline       $0$	                     &  $1$                                    \\			
			\hline   $(p-1)p^{K-2}$	                     &      $\frac{p+1}{2}p^{K-2v-1}+\frac{p-1}{2}{L^{m_1}}p^{\frac{K-2}{2}-v}-1$  \\   
			\hline   $(p-1)\big(p^{K-2}-{L^{m_1}}p^{\frac{K-4}{2}+v}\big)$   &  $p^{K}-p^{K-2v}$        \\ 
			
			\hline   $(p-1)p^{K-2}-2{L^{m_1}}p^{\frac{K-2}{2}+v}$   &  $\frac{p-1}{2}\big(p^{K-2v-1}-{L^{m_1}}p^{\frac{K-2}{2}-v}\big)$        \\ 
			\hline
		\end{tabular} 
	\end{center}  Furthermore, $\mathcal{C}_{\tilde{D}_\lambda}$ is a $\big{[}\frac{1}{2}\big(p^{K-1}-{L^{m_1}}p^{\frac{K-2}{2}+v}\big),K\big{]}$ code with weight enumerator in Table $10^{\circ}$.
\begin{center} Table $10^{\circ}$~~~ The weight enumerator of  $\mathcal{C}_{\tilde{D}_\lambda}$.
	
	\begin{tabular}{|p{5cm}<{\centering}| p{6cm}<{\centering}|}
		\hline   weight $w$	                     &   frequency $A_w$                    \\ 
		\hline       $0$	                     &  $1$                                    \\			
		\hline   $\frac{p-1}{2}p^{K-2}$	                     &      $\frac{p+1}{2}p^{K-2v-1}+\frac{p-1}{2}{L^{m_1}}p^{\frac{K-2}{2}-v}-1$  \\   
		\hline   $\frac{p-1}{2}\big(p^{K-2}-{L^{m_1}}p^{\frac{K-4}{2}+v}\big)$   &  $p^{K}-p^{K-2v}$        \\ 
		
		\hline   $\frac{p-1}{2}p^{K-2}-{L^{m_1}}p^{\frac{K-2}{2}+v}$   &  $\frac{p-1}{2}\big(p^{K-2v-1}-{L^{m_1}}p^{\frac{K-2}{2}-v}\big)$        \\ 
		\hline
	\end{tabular} 
\end{center} 
\end{theorem}
\begin{theorem}\label{t11}
	For $\lambda\in \mathbb{F}_p^{*}$, if  $\frac{m_2}{v}\equiv 0~(\mathrm{mod}~4)$ and $m_1$ is odd, then $\mathcal{C}_{D_\lambda}$ is a $\big{[}p^{K-1}-\eta_{1}(-\lambda)L^{m_1+1}p^{\frac{K-1}{2}+v}, K\big{]}$ code with weight enumerator in Table $11$, and  the complete weight enumerator is 
	\begin{align}\label{w11}{\scriptsize
		\begin{aligned}
		&\mathrm{W}(\mathcal{C}_{D_\lambda})\\=
		&w_0^{p^{K-1}-\eta_{1}(-\lambda)L^{m_1+1}p^{\frac{K-1}{2}+v}}+\big(p^{K-2v-1}-1\big)w_0^{p^{K-2}-\eta_{1}(-\lambda)L^{m_1+1}p^{\frac{K-1}{2}}}\prod_{i\in \mathbb{F}_p^*}w_i^{p^{K-2}} \\&+\big(p^{K}-\frac{p+1}{2}p^{K-2v-1}+\frac{p-1}{2}\eta_{1}(-\lambda)L^{m_1+1}p^{\frac{K-3}{2}-v}\big)\prod_{i\in \mathbb{F}_p}w_i^{p^{K-2}-\eta_{1}(-\lambda)L^{m_1+1}p^{\frac{K-3}{2}}}
		\\&+\big(p^{K-2v-1}-\eta_{1}(-\lambda){L^{m_1+1}}p^{\frac{K-1}{2}-v}\big)\!\!\!\!\!\sum_{\substack{j\in\mathrm{F}_p^{*}\\ \eta_1(j)=\eta_1(\lambda)}}\!\!\!\!\!
		w_{i=\pm\sqrt{4\lambda j}}^{p^{K-2}-\eta_{1}(-\lambda){{L^{m_1+1}}}(p-1)p^{\frac{K-1}{2}+v}}\!\!\!\!\!\!\!\prod_{\substack{i\in \mathbb{F}_p\\i\neq\pm\sqrt{4\lambda j}}}w_i^{p^{K-2}+\eta_{1}(-\lambda){{L^{m_1+1}}}p^{\frac{K-3}{2}+v}}.  
		\end{aligned}}
	\end{align}	
	
	\begin{center} Table $11$~~~ The weight enumerator of  $\mathcal{C}_{D_\lambda}$
		
		\begin{tabular}{|p{7.5cm}<{\centering}| p{7.5cm}<{\centering}|}
			\hline   weight $w$	                     &   frequency $A_w$                    \\ 
			\hline       $0$	                     &  $1$                                    \\ 
			\hline   $(p-1)p^{K-2}$   & $p^{K-2v-1}-1$       \\ 
			\hline   $(p-1)(p^{K-2}-\eta_{1}(-\lambda)L^{m_1+1}p^{\frac{K-3}{2}+v})$	                     &    $p^{K}-\frac{p+1}{2}p^{K-2v-1}+\frac{p-1}{2}\eta_{1}(-\lambda)L^{m_1+1}p^{\frac{K-1}{2}-v}$ \\
			\hline   $(p-1)p^{K-2}-(p+1)\eta_{1}(-\lambda){L^{m_1+1}}p^{\frac{K-3}{2}+v}$   &  $\frac{p-1}{2}\big(p^{K-2v-1}-\eta_{1}(-\lambda){L^{m_1+1}}p^{\frac{K-1}{2}-v}\big)$        \\        
			\hline
		\end{tabular} 
	\end{center}  Furthermore, $\mathcal{C}_{\tilde{D}_\lambda}$ is a $\big{[}\frac{1}{2}\big(p^{K-1}-\eta_{1}(-\lambda)L^{m_1+1}p^{\frac{K-1}{2}+v}\big), K\big{]}$ code with weight enumerator in Table $11^{\circ}$.
	
	\begin{center} Table $11^{\circ}$~~~ The weight enumerator of  $\mathcal{C}_{D_0}$
		
		\begin{tabular}{|p{7.7cm}<{\centering}| p{7.5cm}<{\centering}|}
			\hline   weight $w$	                     &   frequency $A_w$                    \\ 
			\hline       $0$	                     &  $1$                                    \\ 
			\hline   $\frac{p-1}{2}p^{K-2}$   & $p^{K-2v-1}-1$       \\ 
			\hline   $\frac{p-1}{2}(p^{K-2}-\eta_{1}(-\lambda)L^{m_1+1}p^{\frac{K-3}{2}+v})$	                     &    $p^{K}-\frac{p+1}{2}p^{K-2v-1}+\frac{p-1}{2}\eta_{1}(-\lambda)L^{m_1+1}p^{\frac{K-1}{2}-v}$ \\
			\hline   $\frac{1}{2}\big((p-1)p^{K-2}-(p+1)\eta_{1}(-\lambda){L^{m_1+1}}p^{\frac{K-3}{2}+v}\big)$   &  $\frac{p-1}{2}\big(p^{K-2v-1}-\eta_{1}(-\lambda){L^{m_1+1}}p^{\frac{K-1}{2}-v}\big)$        \\        
			\hline
		\end{tabular} 
	\end{center} 
\end{theorem}

\section{Proofs of main results}
\subsection{Some auxiliary lemmas}

Lemmas \ref{l31}-\ref{l32}  are useful for calculating the length and the weights for $\mathcal{C}_{D_\lambda}$.	

\begin{lemma}\label{l31}
	For $\lambda\in \mathbb{F}_p$ and
	\begin{align*}\small
	N_\lambda=\# \big{\{}(x,y)\in \mathbb{F}_{p^{m_1}}\times\mathbb{F}_{p^{m_2}}\backslash\{(0,0)\}~|~\mathrm{Tr_{m_1}}(x^2)+\mathrm{Tr_{m_2}}(y^{p^{u+1}})=\lambda~\big{\}},
	\end{align*} the following assertions hold.
	
	$(1)$ For $\lambda=0$,
	\begin{align}\label{n1}
	N_\lambda=
	&\begin{cases}
	p^{K-1}-1,\quad &\text{if}~K~\text{is odd};\\ 
	p^{K-1}+(p-1)L^{K}p^{\frac{K-2}{2}}-1,\quad &\text{if}~\frac{m_2}{v}~\text{is odd}~\text{and}~K~\text{is even};\\ 
	p^{K-1}+(p-1)L^{m_1}p^{\frac{K-2}{2}}-1,\quad &\text{if}~\frac{m_2}{v}\equiv 2~(\mathrm{mod}~4)~\text{and}~m_1~\text{is even};\\     
	p^{K-1}+(p-1)L^{m_1}p^{\frac{K-2}{2}+v}-1,\quad &\text{if}~\frac{m_2}{v}\equiv 0~(\mathrm{mod}~4)~\text{and}~m_1~\text{is even}.       		
	\end{cases} 	
	\end{align} 
	
	$(2)$ For $\lambda\neq 0$,	
	\begin{align}\label{nlambda}
	N_\lambda
	=&\begin{cases}
	p^{K-1}-\eta_{1}(-\lambda){{L^{K+1}}}p^{\frac{K-1}{2}},\quad &\text{if}~\frac{m_2}{v}~\text{and}~K~\text{are both odd};\\ 
	p^{K-1}-{L^{K}}p^{\frac{K-2}{2}},\quad &\text{if}~\frac{m_2}{v}~\text{is odd}~\text{and}~K~\text{is even};\\ 
	p^{K-1}-\eta_{1}(-\lambda){{L^{m_1+1}}}p^{\frac{K-1}{2}},\quad &\text{if}~\frac{m_2}{v}\equiv 2~(\mathrm{mod}~4)~\text{and}~m_1~\text{is odd};\\     
	p^{K-1}-{L^{m_1}}p^{\frac{K-2}{2}},\quad &\text{if}~\frac{m_2}{v}\equiv 2~(\mathrm{mod}~4)~\text{and}~m_1~\text{is even};\\         
	p^{K-1}-{L^{m_1}}p^{\frac{K-2}{2}+v},\quad &\text{if}~\frac{m_2}{v}\equiv 0~(\mathrm{mod}~4)~\text{and}~m_1~\text{is odd};\\     
	p^{K-1}-\eta_{1}(-\lambda)L^{m_1+1}p^{\frac{K-1}{2}},\quad &\text{if}~\frac{m_2}{v}\equiv 0~(\mathrm{mod}~4)~\text{and}~m_1~\text{is even}.\\         		
	\end{cases}
	\end{align}
\end{lemma}
{\bf Proof}. It follows from Lemma $\ref{lS2}$ that
\begin{align*}
N
=& \sum_{x\in\mathbb{F}_{p^{m_1}}}\sum_{y\in\mathbb{F}_{p^{m_2}}}
\bigg(p^{-1}\sum_{z_1\in\mathbb{F}_p}\zeta_p^{z_1(\mathrm{Tr}_{m_1}(x^2)+\mathrm{Tr}_{m_2}(y^{p^u+1})-\lambda)}\bigg)\\
=&p^{K-1}-1+p^{-1}\sum_{z_1\in\mathbb{F}_p^*}\zeta_p^{-\lambda z_1}\sum_{x\in\mathbb{F}_{p^{m_1}}}\zeta_p^{\mathrm{Tr}_{m_1}(z_1x^2)}
\sum_{y\in\mathbb{F}_{p^{m_2}}}\zeta_p^{\mathrm{Tr}_{m_2}(z_1y^{p^u+1})}\\
=&p^{K-1}+p^{-1}\sum_{z_1\in \mathrm{F}_{p}^*}\zeta_p^{-\lambda z_1}Q_{m_1}(z_1,0)S_{m_2,u}(z_1,0)\\
=&p^{K-1}+p^{-1}G_{m_1}\sum_{z_1\in\mathrm{F}_{p}^{*}}\zeta_p^{-\lambda z_1}\eta_{m_1}(z_1)S_{m_2,u}(z_1,0)\\
=&\begin{cases}
p^{K-1}+p^{-1}G_{m_1}G_{m_2}\sum_{z_1\in\mathrm{F}_{p}^{*}}\zeta_p^{-\lambda z_1}\eta_{m_1}(z_1)\eta_{m_2}(z_1),\quad &\text{if}~\frac{m_2}{v}~\text{is odd};\\ 
p^{K-1}-p^{-1}G_{m_1}p^{s}\sum_{z_1\in\mathrm{F}_{p}^{*}}\zeta_p^{-\lambda z_1}\eta_{m_1}(z_1),\quad &\text{if}~\frac{m_2}{v}\equiv 2~(\mathrm{mod}~4);\\
p^{K-1}-p^{-1}G_{m_1}p^{s+v}\sum_{z_1\in\mathrm{F}_{p}^{*}}\zeta_p^{-\lambda z_1}\eta_{m_1}(z_1),\quad &\text{if}~\frac{m_2}{v}\equiv 0~(\mathrm{mod}~4).\\        		
\end{cases} 
\end{align*}   
If $\lambda=0$, then
\begin{align}\label{NLE0}
\begin{aligned}
N_\lambda=&N-1\\= &\begin{cases}
p^{K-1}-1+p^{-1}G_{m_1}G_{m_2}\sum_{z_1\in\mathrm{F}_{p}^{*}}\eta_{m_1}(z_1)\eta_{m_2}(z_1),\quad &\text{if}~\frac{m_2}{v}~\text{is odd};\\ 
p^{K-1}-1-p^{-1}G_{m_1}p^{s}\sum_{z_1\in\mathrm{F}_{p}^{*}}\eta_{m_1}(z_1),\quad &\text{if}~\frac{m_2}{v}\equiv 2~(\mathrm{mod}~4);\\
p^{K-1}-1-p^{-1}G_{m_1}p^{s+v}\sum_{z_1\in\mathrm{F}_{p}^{*}}\eta_{m_1}(z_1),\quad &\text{if}~\frac{m_2}{v}\equiv 0~(\mathrm{mod}~4);\\        		
\end{cases}\\ =&\begin{cases}
p^{K-1}-1,\quad &\text{if}~K~\text{is odd};\\ 
p^{K-1}+(p-1)p^{-1}G_{m_1}G_{m_2}-1,\quad &\text{if}~\frac{m_2}{v}~\text{is odd}~\text{and}~K~\text{is even};\\ 
p^{K-1}-(p-1)p^{s-1}G_{m_1}-1,\quad &\text{if}~\frac{m_2}{v}\equiv 2~(\mathrm{mod}~4)~\text{and}~m_1~\text{is even};\\     
p^{K-1}-(p-1)p^{s+v-1}G_{m_1}-1,\quad &\text{if}~\frac{m_2}{v}\equiv 0~(\mathrm{mod}~4)~\text{and}~m_1~\text{is even}.       		
\end{cases}
\end{aligned}
\end{align}    
If $\lambda\neq 0$, then
\begin{align}\label{NLN0}\begin{aligned}
&N_\lambda \\	=&\begin{cases}
p^{K-1}+p^{-1}G_{m_1}G_{m_2}\sum_{z_1\in\mathrm{F}_{p}^{*}}\zeta_p^{-\lambda z_1}\eta_{m_1}(z_1)\eta_{1}(z_1),\quad &\text{if}~m_2~\text{is odd};\\     
p^{K-1}+p^{-1}G_{m_1}G_{m_2}\sum_{z_1\in\mathrm{F}_{p}^{*}}\zeta_p^{-\lambda z_1}\eta_{m_1}(z_1),\quad &\text{if}~m_2 \text{~is even and}~\frac{m_2}{v}~\text{is odd};\\ 
p^{K-1}-p^{-1}G_{m_1}p^{s}\sum_{z_1\in\mathrm{F}_{p}^{*}}\zeta_p^{-\lambda z_1}\eta_{m_1}(z_1),\quad &\text{if}~\frac{m_2}{v}\equiv 2~(\mathrm{mod}~4);\\
p^{K-1}-p^{-1}G_{m_1}p^{s+v}\sum_{z_1\in\mathrm{F}_{p}^{*}}\zeta_p^{-\lambda z_1}\eta_{m_1}(z_1),\quad &\text{if}~\frac{m_2}{v}\equiv 0~(\mathrm{mod}~4);\\        		
\end{cases} \\ =&\begin{cases}
p^{K-1}+p^{-1}G_{m_1}G_{m_2}\sum_{z_1\in\mathrm{F}_{p}^{*}}\zeta_p^{-\lambda z_1}\eta_{1}(z_1),\quad &\text{if}~\frac{m_2}{v}~\text{and}~K~\text{are both odd};\\ 
p^{K-1}+p^{-1}G_{m_1}G_{m_2}\sum_{z_1\in\mathrm{F}_{p}^{*}}\zeta_p^{-\lambda z_1} ,\quad &\text{if}~\frac{m_2}{v}~\text{is odd}~\text{and}~K~\text{is even};\\ 
p^{K-1}-p^{-1}G_{m_1}p^{s}\sum_{z_1\in\mathrm{F}_{p}^{*}}\zeta_p^{-\lambda z_1}\eta_{1}(z_1),\quad &\text{if}~\frac{m_2}{v}\equiv 2~(\mathrm{mod}~4)~\text{and}~m_1~\text{is odd};\\     
p^{K-1}-p^{-1}G_{m_1}p^{s}\sum_{z_1\in\mathrm{F}_{p}^{*}}\zeta_p^{-\lambda z_1},\quad &\text{if}~\frac{m_2}{v}\equiv 2~(\mathrm{mod}~4)~\text{and}~m_1~\text{is even};\\         
p^{K-1}-p^{-1}G_{m_1}p^{s+v}\sum_{z_1\in\mathrm{F}_{p}^{*}}\zeta_p^{-\lambda z_1}\eta_{1}(z_1),\quad &\text{if}~\frac{m_2}{v}\equiv 0~(\mathrm{mod}~4)~\text{and}~m_1~\text{is odd};\\     
p^{K-1}-p^{-1}G_{m_1}p^{s+v}\sum_{z_1\in\mathrm{F}_{p}^{*}}\zeta_p^{-\lambda z_1},\quad &\text{if}~\frac{m_2}{v}\equiv 0~(\mathrm{mod}~4)~\text{and}~m_1~\text{is even};\\         		
\end{cases} \\
=&\begin{cases}
p^{K-1}+\eta_{1}(-\lambda)p^{-1}G_1G_{m_1}G_{m_2},\quad \quad &\text{if}~\frac{m_2}{v}~\text{and}~K~\text{are both odd};\\ 
p^{K-1}-p^{-1}G_{m_1}G_{m_2},\quad\quad  &\text{if}~\frac{m_2}{v}~\text{is odd}~\text{and}~K~\text{is even};\\ 
p^{K-1}-\eta_{1}(-\lambda)G_1G_{m_1}p^{s-1},\quad &\text{if}~\frac{m_2}{v}\equiv 2~(\mathrm{mod}~4)~\text{and}~m_1~\text{is odd};\\     
p^{K-1}+G_{m_1}p^{s-1},\quad\quad  &\text{if}~\frac{m_2}{v}\equiv 2~(\mathrm{mod}~4)~\text{and}~m_1~\text{is even};\\         
p^{K-1}-\eta_{1}(-\lambda)G_1G_{m_1}p^{s+v-1},\quad &\text{if}~\frac{m_2}{v}\equiv 0~(\mathrm{mod}~4)~\text{and}~m_1~\text{is odd};\\     
p^{K-1}+G_{m_1}p^{s+v-1},\quad \quad &\text{if}~\frac{m_2}{v}\equiv 0~(\mathrm{mod}~4)~\text{and}~m_1~\text{is even}.\\         		
\end{cases} 
\end{aligned}
\end{align}    

By $(\ref{NLE0})$-$(\ref{NLN0})$ and Lemma \ref{l21}, we complete the proof of Lemma \ref{l31}. $\hfill\Box$\\

\begin{lemma}\label{l32}
	For $\lambda, \rho\in \mathbb{F}_{p}$, $(a,b)\in \mathbb{F}_{q^{m_1}}\times \mathbb{F}_{q^{m_2}}/\{(0,0)\}$ and
	\begin{align*}\small
	& N_{\lambda, \rho}(a,b)\\
	=&\# \big{\{}(x,y)\in \mathbb{F}_{p^{m_1}}\times\mathbb{F}_{p^{m_2}}\big{|}\mathrm{Tr_{m_1}}(x^2)+\mathrm{Tr_{m_2}}(y^{p^{u+1}})=\lambda~\text{and}~\mathrm{Tr_{m_1}}(ax)+
	\mathrm{Tr_{m_2}}(by)=\rho\big{\}},
	\end{align*} 
	denote $\mathrm{T}(a,b)=\mathrm{Tr}_{m_1}\big(\frac{a^2}{4}\big)+\mathrm{Tr}_{m_2}\big(\gamma_b^{p^u+1}\big)$, then the following assertions hold.
	
	$(\mathrm{I})$ For $\lambda=0$,\\
	
	$(1)$ if $\frac{m_2}{v}$ and $K$ are both odd, then
	\begin{align}\label{N1}
	\begin{aligned}
	N_{\lambda,\rho}(a,b)=&
	\begin{cases}
	p^{K-2},\quad&\text{if}~\mathrm{T}(a,b)=0;\\
	p^{K-2}+\eta_1(-\mathrm{T}(a,b))(p-1)L^{K+1}p^{\frac{K-3}{2}},\quad&\text{if}~\mathrm{T}(a,b)\neq0~\text{and}~\rho=0;\\
	p^{K-2}-\eta_1(-\mathrm{T}(a,b))L^{K+1}p^{\frac{K-3}{2}},\quad&\text{if}~\mathrm{T}(a,b)\neq0~\text{and}~\rho\neq0.\\						
	\end{cases}
	\end{aligned}	
	\end{align}
	
	$(2)$ If $\frac{m_2}{v}$ is odd and $K$ is even, then
	\begin{align}\label{N2}
	\begin{aligned}
	N_{\lambda,\rho}(a,b)=&
	\begin{cases}
	p^{K-2}+(p-1)L^{K}p^{\frac{K-2}{2}},\quad&\text{if}~\mathrm{T}(a,b)=0~\text{and}~\rho=0;\\
	p^{K-2},\quad&\text{if}~\mathrm{T}(a,b)=0~\text{and}~\rho\neq0,~\\&~~~\text{or}~\mathrm{T}(a,b)\neq0~\text{and}~\rho=0;\\
	p^{K-2}+L^{K}p^{\frac{K-2}{2}},\quad&\text{if}~\mathrm{T}(a,b)\neq0~\text{and}~\rho\neq0.\\						
	\end{cases}
	\end{aligned}		
	\end{align}
	
	$(3)$ If $\frac{m_2}{v}\equiv 2~(\mathrm{mod}~4)$ and $m_1$ is odd, then
	\begin{align}\label{N3}
	\begin{aligned}		
	N_{\lambda,\rho}(a,b)=&
	\begin{cases}
	p^{K-2},\quad&\text{if}~ \mathrm{T}(a,b)=0;\\
	p^{K-2}-\eta_1\big(-\mathrm{T}(a,b)\big)(p-1)L^{{m_1}+1}p^{\frac{K-3}{2}},\quad&\text{if}~ \mathrm{T}(a,b)\neq0~\text{and}~\rho=0;\\
	p^{K-2}+\eta_1\big(-\mathrm{T}(a,b)\big)L^{{m_1}+1}p^{\frac{K-3}{2}},\quad&\text{if}~ \mathrm{T}(a,b)\neq0~\text{and}~\rho\neq 0.	
	\end{cases}
	\end{aligned}
	\end{align}
	
	$(4)$ If $\frac{m_2}{v}\equiv 2~(\mathrm{mod}~4)$ and $m_1$ is even, then
	\begin{align}\label{N4}
	\begin{aligned}			
	N_{\lambda,\rho}(a,b)=&
	\begin{cases}
	p^{K-2}+(p-1)L^{{m_1}}p^{\frac{K-2}{2}},\quad&\text{if}~ \mathrm{T}(a,b)=0~\text{and}~\rho=0;\\
	p^{K-2}	,\quad&\text{if}~ \mathrm{T}(a,b)\neq0~\text{and}~\rho=0,~\\&~~~\text{or}~ \mathrm{T}(a,b)=0~\text{and}~\rho\neq 0;\\
	p^{K-2}+L^{{m_1}}p^{\frac{K-2}{2}},\quad&\text{if}~ \mathrm{T}(a,b)\neq0~\text{and}~\rho\neq 0.\\	
	\end{cases}
	\end{aligned}
	\end{align}
	
	$(5)$ If $\frac{m_2}{v}\equiv 0~(\mathrm{mod}~4)$ and $m_1$ is odd, then
	{\begin{align}\label{N5}	
		\begin{aligned}	
		&N_{\lambda,\rho}(a,b)\\=&
		\begin{cases}
		p^{K-2},&\!\!\text{if} ~(\ref{E1})~ \text{is not solvable},\\&~~~\text{or} ~(\ref{E1})~\text{is solvable and}~\mathrm{T}(a,b)=0;\\
		p^{K-2}-\eta_1\big(-\mathrm{T}(a,b)\big)(p-1)L^{{m_1}+1}p^{\frac{K-3}{2}+v},&\!\!\text{if}~(\ref{E1})~\text{is solvable,}~\mathrm{T}(a,b)\neq0~\text{and}~\rho=0;\\
		p^{K-2}+\eta_1\big{(}-\mathrm{T}(a,b)\big)L^{{m_1}+1}p^{\frac{K-3}{2}+v},&\!\!\text{if}~(\ref{E1})~\text{is solvable,}~ \mathrm{T}(a,b)\neq0~\text{and}~\rho\neq 0.	
		\end{cases}
		\end{aligned}
		\end{align}}
	
	$(6)$ If $\frac{m_2}{v}\equiv 0~(\mathrm{mod}~4)$ and $m_1$ is even, then
	\begin{align}\label{N6}	
	\begin{aligned}	
	N_{\lambda,\rho}(a,b)=&
	\begin{cases}
	p^{K-2}+(p-1)L^{{m_1}}p^{\frac{K-4}{2}+v},\quad\quad &\text{if} ~(\ref{E1})~ \text{is not solvable};\\
	p^{K-2}+(p-1)L^{{m_1}}p^{\frac{K-2}{2}+v},\quad\quad &\text{if}~(\ref{E1})~\text{is solvable,}~\mathrm{T}(a,b)=0~\text{and}~\rho= 0;\\
	p^{K-2}+L^{{m_1}}p^{\frac{K-2}{2}+v},\quad\quad &\text{if}~(\ref{E1})~\text{is solvable,}~\mathrm{T}(a,b)\neq 0~\text{and}~\rho\neq 0;	\\		
	p^{K-2},&\text{if} ~(\ref{E1})~\text{is solvable,}~ \mathrm{T}(a,b)\neq0~\text{and}~\rho=0,\\&~~\text{or} ~(\ref{E1})~\text{is solvable,}~\mathrm{T}(a,b)=0~\text{and}~\rho\neq 0.\\	
	\end{cases}
	\end{aligned}		
	\end{align}

	$(\mathrm{II})$ For $\lambda\neq 0$, \\
	
	$(1)$ if $\frac{m_2}{v}$ and $K$ are odd, then
	{\small\begin{align}\label{N1l}
		\begin{aligned}
		&N_{\lambda,\rho}(a,b)\\=&\begin{cases}	
		p^{K-2}-\eta_{1}(-\lambda)L^{K+1}p^{\frac{K-1}{2}},\quad&\text{if}~~\rho=0\text{~and~} \mathrm{T}(a,b)=0;\\
		p^{K-2},\quad&\text{if}~~\rho\neq0\text{~and~} \mathrm{T}(a,b)=0;\\		
		p^{K-2}-\eta_1\big(-\mathrm{T}(a,b)\big)(p-1)L^{K+1}p^{\frac{K-3}{2}},\quad&\text{if}~\lambda\neq 0,~\mathrm{T}(a,b)\neq0\text{~and~} \rho^2-4\lambda\mathrm{T}(a,b)=0;\\
		p^{K-2}+\eta_1\big(-\mathrm{T}(a,b)\big)L^{K+1}p^{\frac{K-3}{2}},\quad&\text{if}~\lambda\neq 0,~\mathrm{T}(a,b)\neq0\text{~and~} \rho^2-4\lambda\mathrm{T}(a,b)\neq0.
		\end{cases}
		\end{aligned}			
		\end{align}}
	
	$(2)$  If $\frac{m_2}{v}$ is odd and $K$ is even, then
	\begin{align}\label{N2l}
	\begin{aligned}
	&N_{\lambda,\rho}(a,b)\\=&
	\begin{cases}
	p^{K-2}-L^{K}p^{\frac{K-2}{2}},\quad\quad &\text{if}~ \mathrm{T}(a,b)=0~\text{and}~\rho=0;\\
	p^{K-2}	,\quad\quad &\text{if}~\mathrm{T}(a,b)=0~\text{and}~\rho\neq0,\\&~~\text{or}~ \mathrm{T}(a,b)\neq0~\text{and}~\rho^2-4\lambda\mathrm{T}(a,b)= 0;\\
	p^{K-2}+\eta_1(\rho^2-4\lambda\mathrm{T}(a,b))L^{K}p^{\frac{K-2}{2}},\quad\quad &\text{if}~  \mathrm{T}(a,b)\neq0~\text{and}~\rho^2-4\lambda\mathrm{T}(a,b)\neq 0.					
	\end{cases}		
	\end{aligned}		
	\end{align}
	
	$(3)$ If $\frac{m_2}{v}\equiv 2~(\mathrm{mod}~4)$ and $m_1$ is odd, then
	\begin{align}\label{N3l}
	\begin{aligned}		
	&N_{\lambda,\rho}(a,b)\\=&
	\begin{cases}
	p^{K-2}	-\eta_{1}(-\lambda)L^{m_1+1}p^{\frac{K-1}{2}},\quad&\text{if}~~\rho=0\text{~and~} \mathrm{T}(a,b)=0;\\
	p^{K-2},&\text{if}~~\rho\neq0\text{~and~} \mathrm{T}(a,b)=0;\\		
	p^{K-2}-\eta_1\big(-\mathrm{T}(a,b)\big)(p-1)L^{m_1+1}p^{\frac{K-3}{2}},\quad&\text{if}~\mathrm{T}(a,b)\neq0\text{~and~} \rho^2-4\lambda\mathrm{T}(a,b)=0;\\
	p^{K-2}+\eta_1\big(-\mathrm{T}(a,b)\big)L^{m_1+1}p^{\frac{K-3}{2}},\quad&\text{if}~\mathrm{T}(a,b)\neq0\text{~and~} \rho^2-4\lambda\mathrm{T}(a,b)\neq0.
	\end{cases}
	\end{aligned}	
	\end{align}
	
	$(4)$ If $\frac{m_2}{v}\equiv 2~(\mathrm{mod}~4)$ and $m_1$ is even, then
	\begin{align}\label{N4l}
	\begin{aligned}			
	&N_{\lambda,\rho}(a,b)\\=&
	\begin{cases}
	p^{K-2}+L^{m_1}p^{\frac{K-2}{2}},\quad\quad &\text{if}~ \mathrm{T}(a,b)=0~\text{and}~\rho=0;\\
	p^{K-2}	,\quad\quad &\text{if}~ \mathrm{T}(a,b)=0~\text{and}~\rho\neq0,\\&~~\text{or}~ \mathrm{T}(a,b)\neq0~\text{and}~\rho^2-4\lambda\mathrm{T}(a,b)= 0;\\
	p^{K-2}+\eta_1(\rho^2-4\lambda\mathrm{T}(a,b))L^{m_1}p^{\frac{K-2}{2}},\quad\quad &\text{if}~  \mathrm{T}(a,b)\neq0~\text{and}~\rho^2-4\lambda\mathrm{T}(a,b)\neq 0.	
	\end{cases}
	\end{aligned}
	\end{align}	
	
	$(5)$ If $\frac{m_2}{v}\equiv 0~(\mathrm{mod}~4)$ and $m_1$ is odd, then
	{\small\begin{align}\label{N5l}	
		\begin{aligned}	
		&	N_{\lambda,\rho}(a,b)\\=&
		\begin{cases}
		p^{K-2}	-\eta_{1}(-\lambda)L^{m_1+1}p^{\frac{K-3}{2}+v},\quad&\text{if}~ ~(\ref{E1})~ \text{is not solvable},\\
		p^{K-2}-\eta_{1}(-\lambda)L^{m_1+1}p^{\frac{K-1}{2}+v},\quad&\text{if}~ ~(\ref{E1})~\text{is solvable},~\mathrm{T}(a,b)=0\text{~and~}\rho=0;\\
		p^{K-2},\quad&\text{if}~ ~(\ref{E1})~\text{is solvable},~\mathrm{T}(a,b)=0\text{~and~}\rho\neq0;\\	
		p^{K-2}-\eta_1\big(-\mathrm{T}(a,b)\big)(p-1)L^{m_1+1}p^{\frac{K-3}{2}+v},\quad&\text{if}~ ~(\ref{E1})~\text{is solvable},~\mathrm{T}(a,b)\neq0\\&~~\text{~and~}\rho^2-4\lambda\mathrm{T}(a,b)=0;\\
		p^{K-2}+\eta_1\big(-\mathrm{T}(a,b)\big)L^{m_1+1}p^{\frac{K-3}{2}+v},\quad&\text{if}~ ~(\ref{E1})~\text{is solvable},~\mathrm{T}(a,b)\neq0\\&~~\text{~and~}\rho^2-4\lambda\mathrm{T}(a,b)\neq0.	
		\end{cases}
		\end{aligned}
		\end{align}}
	
	$(6)$ If $\frac{m_2}{v}\equiv 0~(\mathrm{mod}~4)$ and $m_1$ is even, then
	\begin{align}\label{N6l}	
	\begin{aligned}	
	&	N_{\lambda,\rho}(a,b)\\=&
	\begin{cases}
	p^{K-2}-L^{m_1}p^{\frac{K-4}{2}+v},\quad&\text{if}~(\ref{E1})~ \text{is not solvable};\\
	p^{K-2}-L^{m_1}p^{\frac{K-2}{2}+v},\quad&\text{if}~(\ref{E1})~\text{is solvable,}~\mathrm{T}(a,b)=0~\text{and}~\rho=0;\\		
	p^{K-2},\quad&\text{if}~~(\ref{E1})~\text{is solvable,}~\mathrm{T}(a,b)=0~\text{and}~\rho\neq 0;\\
	&~~\text{or}~(\ref{E1})~\text{is solvable,}~\mathrm{T}(a,b)\neq0~\\&~~\text{and}~\rho^2-4\lambda\mathrm{T}(a,b)=0;\\
	p^{K-2}+\eta_1(\rho^2-4\lambda\mathrm{T}(a,b))L^{m_1}p^{\frac{K-2}{2}+v},\quad&\text{if}~(\ref{E1})~\text{is solvable,}~~\mathrm{T}(a,b)\neq0\\&~~\text{~and~}\rho^2-4\lambda\mathrm{T}(a,b)\neq0.
	\end{cases}
	\end{aligned}		
	\end{align}
\end{lemma}
{\bf Proof}. By calculating directly, we have
\begin{align*}
&N_{\lambda,\rho}(a,b)\\
=& \sum_{x\in\mathbb{F}_{p^{m_1}}}\sum_{y\in\mathbb{F}_{p^{m_2}}}
\bigg(p^{-1}\sum_{z_1\in\mathbb{F}_p}\zeta_p^{z_1(\mathrm{Tr}_{m_1}(x^2)+\mathrm{Tr}_{m_2}(y^{p^u+1})-\lambda)}\bigg)
\bigg(p^{-1}\sum_{z_2\in\mathbb{F}_p}\zeta_p^{z_2(\mathrm{Tr}_{m_1}(ax)+\mathrm{Tr}_{m_2}(by)-\rho)}\bigg)\\
=& p^{K-2}+p^{-2}\sum_{z_2\in\mathbb{F}_p^*}\zeta_p^{-\rho z_2}\sum_{x\in\mathbb{F}_{p^{m_1}}}
\zeta_p^{\mathrm{Tr}_{m_1}(z_2ax)}\sum_{y\in\mathbb{F}_{p^{m_2}}}\zeta_p^{\mathrm{Tr}_{m_2}(z_2by)}\\
&+ p^{-2}\sum_{z_2\in\mathbb{F}_p}\zeta_p^{-\rho z_2}\sum_{z_1\in\mathbb{F}_p^*}\zeta_p^{-\lambda z_1}
\sum_{x\in\mathbb{F}_{p^{m_1}}}\zeta_p^{\mathrm{Tr}_{m_1}(z_1x^2 +z_2ax)}\sum_{y\in\mathbb{F}_{p^{m_2}}}\zeta_p^{\mathrm{Tr}_{m_2}(z_1y^{p^u+1}+z_2by)}\\
=& p^{K-2}+p^{-2}\Omega,	
\end{align*}
where

\begin{align*}
\Omega=\sum_{z_1\in \mathrm{F}_{p}^{*}}\zeta_p^{-\lambda z_1}\sum_{z_2\in \mathrm{F}_{p}}\zeta_p^{-\rho z_2}Q_{m_1}(z_1,z_2a)S_{m_2,u}(z_1,z_2b).
\end{align*}
Now by Lemmas $\ref{l21}$ and $\ref{lS1}$-$\ref{lS2}$, we calculate $\Omega$ as follows.\\

{ Case 1}. For  odd $\frac{m_2}{v}$, 

\begin{align*}
\Omega
=&\sum_{z_1\in \mathrm{F}_{p}^*}\zeta_p^{-\lambda z_1}\sum_{z_2\in \mathrm{F}_{p}}\zeta_p^{-\rho z_2}\big(G_{m_1}\zeta_p^{\mathrm{Tr}_{m_1}(-\frac{z_2^2a^2}{4z_1})}\eta_{m_1}(z_1)G_{m_2}\zeta_p^{\mathrm{Tr}_{m_2}
	(-z_1(\frac{z_2}{z_1}\gamma_b)^{p^u+1})}\eta_{m_2}(z_1)\big)\\
=& G_{m_1}G_{m_2}\bigg(\sum_{z_1\in \mathrm{F}_{p}^*}\zeta_p^{-\lambda z_1}\sum_{z_2\in \mathrm{F}_{p}}\zeta_p^{-\rho z_2}\eta_{m_1}(z_1)\eta_{m_2}(z_1)\zeta_p^{
	-z_1\big(\mathrm{Tr}_{m_1}((\frac{z_2}{z_1})^2\frac{a^2}{4})+\mathrm{Tr}_{m_2}((\frac{z_2}{z_1})^{2}\gamma_b^{p^u+1})\big)}   \bigg) \\
=& G_{m_1}G_{m_2}\bigg(
\sum_{z_1\in \mathrm{F}_{p}^{*}}\zeta_p^{-\lambda z_1}\eta_{m_1}(z_1)\eta_{m_2}(z_1)\sum_{z_3\in \mathrm{F}_{p}}\zeta_p^{-\mathrm{T}(a,b)z_1z_3^2-\rho z_1 z_3}   \bigg).     
\end{align*}
	
If $K$ is odd, then
\begin{align*}
&\Omega\\
=& G_{m_1}G_{m_2}\big(\sum_{z_1\in \mathrm{F}_{p}^*}\zeta_p^{-\lambda z_1}\eta_{1}(z_1)\sum_{z_3\in \mathrm{F}_{p}}\zeta_p^{
	-\mathrm{T}(a,b)z_1z_3^2-\rho z_1 z_3}   \big)  \\
=&\begin{cases}
G_{m_1}G_{m_2}\big(\sum_{z_1\in \mathrm{F}_{p}^*}\zeta_p^{-\lambda z_1}\eta_{1}(z_1)\sum_{z_3\in \mathrm{F}_{p}}\zeta_p^{-\rho z_1 z_3}   \big) ,\quad\quad\quad&\text{if}~ \mathrm{T}(a,b)=0;\\
\eta_1\big(-\mathrm{T}(a,b)\big)G_1	G_{m_1}G_{m_2}\sum_{z_1\in \mathrm{F}_{p}^*}\zeta_p^{\frac{\rho^2-4\mathrm{T}(a,b)\lambda}{4\mathrm{T}(a,b)}z_1},\quad\quad\quad&\text{if}~ \mathrm{T}(a,b)\neq0;
\end{cases}\\
=&\begin{cases}
0,\quad&\text{if}~\lambda=0\text{~and~} \mathrm{T}(a,b)=0;\\
\eta_1\big(-\mathrm{T}(a,b)\big)(p-1)G_1G_{m_1}G_{m_2},\quad\quad&\text{if}~\lambda=0,~\rho=0\text{~and~}  \mathrm{T}(a,b)\neq0;\\
-\eta_1\big(-\mathrm{T}(a,b)\big)G_1G_{m_1}G_{m_2},\quad\quad&\text{if}~\lambda=0,~\rho\neq 0\text{~and~}  \mathrm{T}(a,b)\neq0;\\	
\eta_{1}(-\lambda)pG_1G_{m_1}G_{m_2},\quad\quad&\text{if}~\lambda\neq0,~\rho=0\text{~and~} \mathrm{T}(a,b)=0;\\
0,\quad\quad&\text{if}~\lambda\neq0,~\rho\neq0\text{~and~} \mathrm{T}(a,b)=0;\\		
\eta_1\big(-\mathrm{T}(a,b)\big)(p-1)G_1G_{m_1}G_{m_2},\quad\quad&\text{if}~\lambda\neq 0,~\mathrm{T}(a,b)\neq0\text{~and~} \rho^2-4\lambda\mathrm{T}(a,b)=0;\\
-\eta_1\big(-\mathrm{T}(a,b)\big)G_1G_{m_1}G_{m_2},\quad\quad&\text{if}~\lambda\neq 0,~\mathrm{T}(a,b)\neq0\text{~and~} \rho^2-4\lambda\mathrm{T}(a,b)\neq0.\\
\end{cases}
\end{align*}	

If $K$ is even, then
\begin{align*}
&\Omega\\=&
G_{m_1}G_{m_2}\bigg(\sum_{z_1\in \mathrm{F}_{p}^*}\zeta_p^{-\lambda z_1}\sum_{z_3\in \mathrm{F}_{p}}\zeta_p^{
	-\mathrm{T}(a,b)z_1z_3^2-\rho z_1z_3}   \bigg) \\
=&\begin{cases}
G_{m_1}G_{m_2}\big(\sum_{z_1\in \mathrm{F}_{p}^*}\zeta_p^{-\lambda z_1}\sum_{z_3\in \mathrm{F}_{p}}\zeta_p^{-\rho z_1 z_3}   \big) ,\quad\quad&\text{if}~ \mathrm{T}(a,b)=0;\\
\eta_1(-1)G_1	G_{m_1}G_{m_2}\sum_{z_1\in \mathrm{F}_{p}^*}\zeta_p^{-\lambda z_1}\eta_1\big(\frac{z_1}{4\mathrm{T}(a,b)}\big)\zeta_p^{\frac{\rho^2}{4\mathrm{T}(a,b)}z_1},\quad\quad&\text{if}~ \mathrm{T}(a,b)\neq0;
\end{cases}\\	
=&\begin{cases}
p(p-1)G_{m_1}G_{m_2},\quad\quad&\text{if}~\lambda=0, \mathrm{T}(a,b)=0~\text{and}~\rho=0;\\
0	,\quad\quad&\text{if}~\lambda=0 ,\mathrm{T}(a,b)=0~\text{and}~\rho\neq 0,\\&~~\text{or}~\lambda=0 , \mathrm{T}(a,b)\neq0~\text{and}~\rho=0 ;\\
\eta_1(-1)G_1^2G_{m_1}G_{m_2},\quad\quad&\text{if}~\lambda=0,  \mathrm{T}(a,b)\neq0~\text{and}~\rho\neq 0;\\
-pG_{m_1}G_{m_2},\quad\quad&\text{if}~\lambda\neq0, \mathrm{T}(a,b)=0~\text{and}~\rho=0;\\
0	,\quad\quad&\text{if}~\lambda\neq0 , \mathrm{T}(a,b)=0~\text{and}~\rho\neq0,\\&~~\text{or}~ \lambda\neq0 ,\mathrm{T}(a,b)\neq0~\text{and}~\rho^2-4\lambda\mathrm{T}(a,b)= 0;\\
\eta_1(4\lambda\mathrm{T}(a,b)-\rho^2)G_1^2G_{m_1}G_{m_2},\quad\quad&\text{if}~\lambda\neq0,  \mathrm{T}(a,b)\neq0~\text{and}~\rho^2-4\lambda\mathrm{T}(a,b)= 0.\\			
\end{cases}
\end{align*}

{ Case 2}. For $\frac{m_2}{v}\equiv 2~(\mathrm{mod}~4)$, 		
\begin{align*}
 \Omega=&\sum_{z_1\in \mathrm{F}_{p}^*}\zeta_p^{-\lambda z_1}\sum_{z_2\in \mathrm{F}_{p}}\zeta_p^{-\rho z_2}Q_{m_1}(z_1,z_2a)S_{m_2,u}(z_1,z_2b)\\
=&-p^s\sum_{z_1\in \mathrm{F}_{p}^*}\zeta_p^{-\lambda z_1}\sum_{z_2\in \mathrm{F}_{p}}\zeta_p^{-\rho z_2}\bigg(G_{m_1}\zeta_p^{\mathrm{Tr}_{m_1}(-\frac{z_2^2a^2}{z_1})}\eta_{m_1}(z_1)\zeta_p^{\mathrm{Tr}_{m_2}
	(-z_1(\frac{z_2}{z_1}\gamma_b)^{p^u+1}}\bigg)\\
=&- p^sG_{m_1}\sum_{z_1\in \mathrm{F}_{p}^*}\zeta_p^{-\lambda z_1}\eta_{m_1}(z_1)
\sum_{z_2\in \mathrm{F}_{p}}\zeta_p^{-\rho z_2}\zeta_p^{
	-z_1\big(\mathrm{Tr}_{m_1}((\frac{z_2}{z_1})^2a^2)+\mathrm{Tr}_{m_2}((\frac{z_2}{z_1}\gamma_b)^{p^u+1})\big)} \\
=&- p^sG_{m_1}
\sum_{z_1\in \mathrm{F}_{p}^*}\zeta_p^{-\lambda z_1}\eta_{m_1}(z_1)\sum_{z_3\in \mathrm{F}_{p}}\zeta_p^{
	-\mathrm{T}(a,b)z_1z_3^2-\rho z_1z_3} .	    
\end{align*}

If $m_1$ is odd, then
\begin{align*}
\Omega
=&-p^s G_{m_1}\big(\sum_{z_1\in \mathrm{F}_{p}^*}\zeta_p^{-\lambda z_1}\eta_{1}(z_1)\sum_{z_3\in \mathrm{F}_{p}}\zeta_p^{
	-\mathrm{T}(a,b)z_1z_3^2-\rho z_1 z_3}   \big)  \\
=&\begin{cases}
-p^sG_{m_1}\big(\sum_{z_1\in \mathrm{F}_{p}^*}\zeta_p^{-\lambda z_1}\eta_{1}(z_1)\sum_{z_3\in \mathrm{F}_{p}}\zeta_p^{-\rho z_1 z_3}   \big) ,\quad\quad\quad&\text{if}~ \mathrm{T}(a,b)=0;\\
-\eta_1\big(-\mathrm{T}(a,b)\big)p^sG_1	G_{m_1}\sum_{z_1\in \mathrm{F}_{p}^*}\zeta_p^{-\lambda z_1}\zeta_p^{\frac{\rho^2}{4\mathrm{T}(a,b)}z_1},\quad\quad\quad&\text{if}~ \mathrm{T}(a,b)\neq0;
\end{cases}\\
=&\begin{cases}
0,\quad&\text{if}~\lambda=0\text{~and~} \mathrm{T}(a,b)=0;\\
-\eta_1\big(-\mathrm{T}(a,b)\big)(p-1)p^sG_1G_{m_1},\quad\quad&\text{if}~\lambda=0,~\rho=0\text{~and~}  \mathrm{T}(a,b)\neq0;\\
\eta_1\big(-\mathrm{T}(a,b)\big)p^sG_1G_{m_1},\quad\quad&\text{if}~\lambda=0,~\rho\neq 0\text{~and~}  \mathrm{T}(a,b)\neq0;\\	
-\eta_{1}(-\lambda)p^{s+1}G_1G_{m_1},\quad\quad&\text{if}~\lambda\neq0,~\rho=0\text{~and~} \mathrm{T}(a,b)=0;\\
0,\quad\quad&\text{if}~\lambda\neq0,~\rho\neq0\text{~and~} \mathrm{T}(a,b)=0;\\		
-\eta_1\big(-\mathrm{T}(a,b)\big)(p-1)p^sG_1G_{m_1},\quad\quad&\text{if}~\lambda\neq 0,~\mathrm{T}(a,b)\neq0\text{~and~} \rho^2-4\lambda\mathrm{T}(a,b)=0;\\
\eta_1\big(-\mathrm{T}(a,b)\big)p^sG_1G_{m_1},\quad\quad&\text{if}~\lambda\neq 0,~\mathrm{T}(a,b)\neq0\text{~and~} \rho^2-4\lambda\mathrm{T}(a,b)\neq0.\\
\end{cases}
\end{align*}

If $m_1$ is even, then
\begin{align*}
\Omega=&
-p^sG_{m_1}\bigg(\sum_{z_1\in \mathrm{F}_{p}^*}\zeta_p^{-\lambda z_1}\sum_{z_3\in \mathrm{F}_{p}}\zeta_p^{
	-\mathrm{T}(a,b)z_1z_3^2-\rho z_1z_3}   \bigg) \\
=&\begin{cases}
-p^sG_{m_1}\big(\sum_{z_1\in \mathrm{F}_{p}^*}\zeta_p^{-\lambda z_1}\sum_{z_3\in \mathrm{F}_{p}}\zeta_p^{-\rho z_1 z_3}   \big) ,\quad\quad&\text{if}~ \mathrm{T}(a,b)=0;\\
-\eta_1(-1)p^sG_1G_{m_1}\sum_{z_1\in \mathrm{F}_{p}^*}\zeta_p^{-\lambda z_1}\eta_1\big(\frac{z_1}{4\mathrm{T}(a,b)}\big)\zeta_p^{\frac{\rho^2}{4\mathrm{T}(a,b)}z_1},\quad\quad&\text{if}~ \mathrm{T}(a,b)\neq0;
\end{cases}\\	
=&\begin{cases}
-(p-1)p^{s+1}G_{m_1},\quad\quad&\text{if}~\lambda=0, \mathrm{T}(a,b)=0~\text{and}~\rho=0;\\
0	,\quad\quad&\text{if}~\lambda=0 ,\mathrm{T}(a,b)=0~\text{and}~\rho\neq 0,\\&~~\text{or}~\lambda=0 , \mathrm{T}(a,b)\neq0~\text{and}~\rho=0 ;\\
-\eta_1(-1)p^sG_1^2G_{m_1},\quad\quad&\text{if}~\lambda=0,  \mathrm{T}(a,b)\neq0~\text{and}~\rho\neq 0;\\
-p^{s+1}G_{m_1},\quad\quad&\text{if}~\lambda\neq0, \mathrm{T}(a,b)=0~\text{and}~\rho=0;\\
0	,\quad\quad&\text{if}~\lambda\neq0 , \mathrm{T}(a,b)=0~\text{and}~\rho\neq0,\\&~~\text{or}~ \lambda\neq0 ,\mathrm{T}(a,b)\neq0~\text{and}~\rho^2-4\lambda\mathrm{T}(a,b)= 0;\\
-\eta_1(4\lambda\mathrm{T}(a,b)-\rho^2)p^sG_1^2G_{m_1},\quad\quad&\text{if}~\lambda\neq0,  \mathrm{T}(a,b)\neq0~\text{and}~\rho^2-4\lambda\mathrm{T}(a,b)\neq 0.\\			
\end{cases}
\end{align*}

{ Case 3}. For $\frac{m_2}{v}\equiv 0~(\mathrm{mod}~4)$,		
\begin{align*}\small
& \Omega\\=&\begin{cases}
- p^{s+v}G_{m_1}\!\!\sum\limits_{z_1\in\mathrm{F}_{p}^{*}}\!\!\!\zeta_p^{-\lambda z_1}\eta_{m_1}(z_1),\qquad\qquad\qquad\qquad\qquad\qquad\qquad\quad\text{if} ~(\ref{E1})~\text{  is not solvable};&\\
- p^{s+v}G_{m_1}\!\!\sum\limits_{z_1\in \mathrm{F}_{p}^*}\!\!\!\zeta_p^{-\lambda z_1}\sum\limits_{z_2\in \mathrm{F}_{p}}\zeta_p^{-\rho z_2}\zeta_p^{\mathrm{Tr}_{m_1}(-\frac{z_2^2a^2}{z_1})}\eta_{m_1}(z_1)\zeta_p^{\mathrm{Tr}_{m_2}
	(-z_1(\frac{z_2}{z_1}\gamma_b)^{p^u+1})},\!~\text{if}~(\ref{E1}) ~ \text{is solvable};&
\end{cases} \\=&\begin{cases}
- p^{s+v}G_{m_1}\!\!\sum\limits_{z_1\in\mathrm{F}_{p}^{*}}\!\!\zeta_p^{-\lambda z_1}\eta_{m_1}(z_1),  &~~\text{if} ~(\ref{E1})~\text{is not solvable};\\
- p^{s+v}G_{m_1}\!\!\sum\limits_{z_1\in \mathrm{F}_{p}^*}\!\!\zeta_p^{-\lambda z_1}\eta_{m_1}(z_1)\sum\limits_{z_3\in \mathrm{F}_{p}}\zeta_p^{-z_1z_3^2\mathrm{T}(a,b)-\rho z_1z_3},&~~\text{if} ~(\ref{E1})~ \text{is solvable}.
\end{cases}   
\end{align*}		

If $m_1$ is odd, then
\begin{align*}
&\Omega
\\
=&\begin{cases}
- p^{s+v}G_{m_1}\sum\limits_{z_1\in\mathrm{F}_{p}^{*}}\zeta_p^{-\lambda z_1}\eta_{1}(z_1),  &\quad\quad \text{if} ~(\ref{E1})~\text{is not solvable};\\
- p^{s+v}G_{m_1}\sum\limits_{z_1\in \mathrm{F}_{p}^*}\zeta_p^{-\lambda z_1}\eta_{1}(z_1)\sum\limits_{z_3\in \mathrm{F}_{p}}\zeta_p^{-z_1\mathrm{T}(a,b)z_3^2-\rho z_1z_3},&\quad\quad \text{if} ~(\ref{E1})~ \text{is solvable};
\end{cases}\\  
=&\begin{cases}
- p^{s+v}G_{m_1}\sum\limits_{z_1\in\mathrm{F}_{p}^{*}}\zeta_p^{-\lambda z_1}\eta_{1}(z_1),\quad\quad  &\text{if} ~(\ref{E1})~\text{is not solvable};\\
-p^{s+v}G_{m_1}\big(\sum\limits_{z_1\in \mathrm{F}_{p}^*}\zeta_p^{-\lambda z_1}\eta_{1}(z_1)\sum\limits_{z_3\in \mathrm{F}_{p}}\zeta_p^{-\rho z_1 z_3}   \big),\quad\quad&\text{if} ~(\ref{E1})~ \text{is solvable and}~\mathrm{T}(a,b)=0;\\
-p^{s+v}\eta_1\big(-\mathrm{T}(a,b)\big)G_1	G_{m_1}\sum\limits_{z_1\in \mathrm{F}_{p}^*}\zeta_p^{\frac{\rho^2-4\lambda \mathrm{T}(a,b)}{4\mathrm{T}(a,b)}z_1},\quad\quad&\text{if} ~(\ref{E1})~ \text{is solvable and}~\mathrm{T}(a,b)\neq0;
\end{cases}\\
=&\begin{cases}
0,\quad&\text{if}~\lambda=0, ~(\ref{E1})~ \text{is not solvable},\\&~~\text{or}~\lambda=0, ~(\ref{E1})~\text{is solvable and}~\mathrm{T}(a,b)=0;\\
-\eta_1\big(-\mathrm{T}(a,b)\big)(p-1)p^{s+v}G_1G_{m_1},\quad&\text{if}~\lambda=0,~(\ref{E1})~\text{is solvable,}~ \mathrm{T}(a,b)\neq0~\text{and}~\rho=0;\\
\eta_1\big(-\mathrm{T}(a,b)\big)p^{s+v}G_1G_{m_1},\quad&\text{if}~\lambda=0,~(\ref{E1})~\text{is solvable,}~ \mathrm{T}(a,b)\neq0~\text{and}~\rho\neq 0;\\
-\eta_{1}(-\lambda)p^{s+v}G_1G_{m_1},\quad&\text{if}~\lambda\neq0, ~(\ref{E1})~ \text{is not solvable};\\
-\eta_{1}(-\lambda)p^{s+v+1}G_1G_{m_1},\quad&\text{if}~\lambda\neq0, ~(\ref{E1})~\text{is solvable},~\mathrm{T}(a,b)=0\text{~and~}\rho=0;\\
0,\quad\quad&\text{if}~\lambda\neq0, ~(\ref{E1})~\text{is solvable}~\mathrm{T}(a,b)=0\text{~and~}\rho\neq0;\\	
-\eta_1\big(-\mathrm{T}(a,b)\big)(p-1)p^{s+v}G_1G_{m_1},\quad&\text{if}~\lambda\neq0, ~(\ref{E1})~\text{is solvable},~\mathrm{T}(a,b)\neq0 \\&~~~\text{~and~}\rho^2-4\lambda\mathrm{T}(a,b)=0;\\
\eta_1\big(-\mathrm{T}(a,b)\big)p^{s+v}G_1G_{m_1},\quad&\text{if}~\lambda\neq0, ~(\ref{E1})~\text{is solvable},~\mathrm{T}(a,b)\neq0\\&~~~\text{~and~}\rho^2-4\lambda\mathrm{T}(a,b)\neq0.	
\end{cases}
\end{align*}

If $m_1$ is even, then
\begin{align*}
&\Omega\\
\\=&\begin{cases}
- p^{s+v}G_{m_1}\!\!\sum\limits_{z_1\in\mathrm{F}_{p}^{*}}\zeta_p^{-\lambda z_1},  &  \quad \text{if} ~(\ref{E1})~\text{is not solvable};\\
- p^{s+v}G_{m_1}\!\!\sum\limits_{z_1\in \mathrm{F}_{p}^*}\zeta_p^{-\lambda z_1}\sum\limits_{z_3\in \mathrm{F}_{p}}\zeta_p^{-z_1z_3^2\mathrm{T}(a,b)-\rho z_1z_3},& \quad \text{if}         ~(\ref{E1})~ \text{is solvable};
\end{cases}\\ 		
=&\begin{cases}
- p^{s+v}G_{m_1}\!\!\sum\limits_{z_1\in\mathrm{F}_{p}^{*}}\zeta_p^{-\lambda z_1}, &\!\!\! \text{if} ~(\ref{E1})~\text{is not solvable};\\
-p^{s+v}G_{m_1}\big(\!\!\sum\limits_{z_1\in \mathrm{F}_{p}^*}\zeta_p^{-\lambda z_1}\sum\limits_{z_3\in \mathrm{F}_{p}}\zeta_p^{-\rho z_1 z_3}  \big),&\!\!\!\text{if} ~(\ref{E1})~ \text{is solvable and}~\mathrm{T}(a,b)=0;\\
-\eta_1\big(-\mathrm{T}(a,b)\big)p^{s+v}G_1	G_{m_1}\!\!\sum\limits_{z_1\in \mathrm{F}_{p}^*}\eta_1(z_1)\zeta_p^{\frac{\rho^2-4\lambda\mathrm{T}(a,b)}{4\mathrm{T}(a,b)}z_1},&\!\!\! \text{if} ~(\ref{E1})~ \text{is solvable and}~\mathrm{T}(a,b)\neq0;
\end{cases}\\
=&\begin{cases}
-(p-1)p^{s+v}G_{m_1},\quad&\text{if}~\lambda=0,(\ref{E1})~ \text{is not solvable};\\
-(p-1)p^{s+v+1}G_{m_1},\quad&\text{if}~\lambda=0,(\ref{E1})~\text{is solvable,}~\mathrm{T}(a,b)=0~\text{and}~\rho=0;\\		
0,\quad&\text{if}~\lambda=0,(\ref{E1})~\text{is solvable,}~\mathrm{T}(a,b)\neq0~\text{and}~\rho=0,\\&~\text{or}~\lambda=0,~(\ref{E1})~\text{is solvable,}~\mathrm{T}(a,b)=0~\text{and}~\rho\neq 0;\\
-\eta_1(-1)p^{s+v}G_1^2G_{m_1},\quad&\text{if}~\lambda=0,(\ref{E1})~\text{is solvable,}~\mathrm{T}(a,b)\neq0~\text{and}~\rho\neq 0.\\	
p^{s+v}G_{m_1},\quad&\text{if}~\lambda\neq0,(\ref{E1})~ \text{is not solvable};\\
p^{s+v+1}G_{m_1},\quad&\text{if}~\lambda\neq0,(\ref{E1})~\text{is solvable,}~\mathrm{T}(a,b)=0~\text{and}~\rho=0;\\		
0,\quad&\text{if}~\lambda\neq0,~(\ref{E1})~\text{is solvable,}~\mathrm{T}(a,b)=0~\text{and}~\rho\neq 0,\\
&~\text{or}~\lambda\neq0,(\ref{E1})~\text{is solvable,}~\mathrm{T}(a,b)\neq0\\&~\text{and}~\rho^2-4\lambda\mathrm{T}(a,b)=0;\\
-\eta_1(4\lambda\mathrm{T}(a,b)-\rho^2)p^{s+v}G_1^2G_{m_1},\quad&\text{if}~\lambda\neq0,(\ref{E1})~\text{is solvable,}~\mathrm{T}(a,b)\neq0\\&~~~\text{~and~}\rho^2-4\lambda\mathrm{T}(a,b)\neq0.	
\end{cases}
\end{align*}

So far, by the { cases} $1$-$3$ and Lemma \ref{l21}, we complete the proof of Lemma \ref{l32}.$\hfill\Box$\\

Lemmas \ref{l35}-\ref{A} are important to calculate the weight enumerators for $\mathcal{C}_{D_\lambda}$.

\begin{lemma}\label{l35}
	For $t\in \mathbb{F}_{p}$ and
	\begin{align*}
	\tilde{A}_{t}=\# \big{\{}(a,b)\in \mathbb{F}_{p^{m_1}}\times\mathbb{F}_{p^{m_2}}|\mathrm{T}(a,b)=t~\big{\}},
	\end{align*}	   		
	if $\frac{m_2}{v}$ is odd  or $\frac{m_2}{v}\equiv 2~(\mathrm{mod}~4)$, then the following assections hold. 
	
	$(1)$ For $t=0$, \begin{align}\label{A3}
	\tilde{A}_{t}=
	&\begin{cases}
	p^{K-1},\quad &\text{if}~K~\text{is odd};\\ 
	p^{K-1}+(p-1)L^{K}p^{\frac{K-2}{2}},\quad &\text{if}~\frac{m_2}{v}~\text{is odd}~\text{and}~K~\text{is even};\\ 
	p^{K-1}+(p-1)L^{m_1}p^{\frac{K-2}{2}},\quad &\text{if}~\frac{m_2}{v}\equiv 2~(\mathrm{mod}~4)~\text{and}~m_1~\text{is even}.
	\end{cases}     			
	\end{align}
	
	$(2)$ For $t\neq0$,
	\begin{align}\label{A4}
	\tilde{A}_t=&\begin{cases}
	p^{K-1}-\eta_{1}(-t)L^{K+1}p^{\frac{K-1}{2}},\quad &\text{if}~\frac{m_2}{v}~\text{and}~K~\text{are odd};\\ 
	p^{K-1}-L^{K}p^{\frac{K-2}{2}},\quad &\text{if}~\frac{m_2}{v}~\text{is odd}~\text{and}~K~\text{is even};\\ 
	p^{K-1}-\eta_{1}(-t)L^{m_1+1}p^{\frac{K-1}{2}},\quad &\text{if}~\frac{m_2}{v}\equiv 2~(\mathrm{mod}~4)~\text{and}~m_1~\text{is odd};\\     
	p^{K-1}-L^{K}p^{\frac{K-2}{2}},\quad &\text{if}~\frac{m_2}{v}\equiv 2~(\mathrm{mod}~4)~\text{and}~m_1~\text{is even}.       		
	\end{cases}
	\end{align}	
\end{lemma}

{\bf Proof}. For odd $\frac{m_2}{v}$, or $\frac{m_2}{v}\equiv 2~(\mathrm{mod}~4)$, it follows from Lemma \ref{l25} that $X^{p^{2u}}+X$ is a permutation polynomial over $\mathbb{F}_{p^{m_2}}[x]$ and then $(\ref{E1})$ has an unique solution in $\mathbb{F}_{p^{m_2}}$, thus, 
\begin{align*}
\tilde{A}_{t}
&=\# \big{\{}(a,b)\in\mathbb{F}_{p^{m_1}}\times\mathbb{F}_{p^{m_2}}|\mathrm{Tr}_{m_1}\big({a^2}\big)+\mathrm{Tr}_{m_2}\big(b^{p^u+1}\big)=t~\big{\}}.
\end{align*}
Now by Lemma \ref{l32}, we can obtain $(\ref{A3})$-$(\ref{A4})$. $\hfill\Box$\\

\begin{lemma}[\cite{GJ2019}, Lemma 13]\label{B}
	For $\frac{m_2}{v}~\equiv~0~(\mathrm{mod}~4)$, and
	\begin{align*}
	B=\big{\{}c\in\mathbb{F}_{p^{m}}~{\big |}~ X^{p^{2u}}+X=c^{p^u}~\text{  is solvable in}~\mathbb{F}_{p^{m}}\big{\}},
	\end{align*} one has
	\begin{align*}
	\# B =p^{m-2v}.
	\end{align*}
\end{lemma}
\begin{lemma}\label{A}
	For $\frac{m_2}{v}~\equiv~0~(\mathrm{mod}~4)$, $t\in\mathbb{F}_p$, and
	\begin{align*}
	\bar{A}_t=\{(a,b)\in\mathbb{F}_{p^{m_1}}\times B~{\big|}~\mathrm{Tr}_{m_1}(\frac{a^2}{4})+\mathrm{Tr}_{m_2}(\gamma_b^{p^u+1})=t\},
	\end{align*} the following two assertions hold.
	
	$(1)$ If $m_1$ is even, then
	\begin{align}\label{m1even}
	\#\bar{A}_t=\begin{cases}
	p^{K-2v-1}+{L^{m_1}}(p-1)p^{\frac{K-2}{2}-v},\qquad&t=0;\\
	p^{K-2v-1}-{L^{m_1}}p^{\frac{K-2}{2}-v},\qquad&\text{otherwise}.\\
	\end{cases}
	\end{align}
	
	$(2)$ If $m_1$ is odd, then
	\begin{align}\label{m1odd}
	\#\bar{A}_t=\begin{cases}
	p^{K-2v-1},\qquad&t=0;\\
	p^{K-2v-1}-\eta_{1}(-t){L^{m_1+1}}p^{\frac{K-1}{2}-v},\qquad&\text{otherwise}.\\
	\end{cases}
	\end{align}
\end{lemma}

To prove Lemma \ref{A}, we need Tables $4$ and $10$, which are given in subsection $4.4$.
\subsection{The proofs for Theorems $\ref{t1}$-$\ref{t3}$ and $\ref{t6}$-$\ref{t9}$}	
{\bf The proofs for Theorems $\ref{t1}$-$\ref{t3}$}.

By Lemmas $\ref{l31}$-$\ref{l35}$, we have the following three cases.

{ Case 1}. If $\frac{m_2}{v}$ and $K$ are both odd, or $\frac{m_2}{v}\equiv 2~(\mathrm{mod}~4)$ and $m_1$ is odd, then the length of $\mathcal{C}_{D_0}$ is $N_0=p^{K-1}-1$. It follows from $(\ref{N1})$ and $(\ref{N3})$ that the nonzero weights of $\mathcal{C}_{D_0}$ are 
\begin{align*}
\begin{aligned}
w_1=(p-1)\big(p^{K-2}-p^{\frac{K-3}{2}}\big),\quad w_2=(p-1)p^{K-2},\quad
w_3=(p-1)\big(p^{K-2}+p^{\frac{K-3}{2}} \big).
\end{aligned}
\end{align*} By Lemma \ref{l35}, we know that $A_{w_2}=\tilde{A}_0-1=p^{K-1}-1$, which combines first two Pless power moments leads to
\begin{align*}
&A_{w_1}=\frac{p-1}{2}p^{\frac{K-1}{2}}\big(p^{\frac{K-1}{2}}+1\big),\quad A_{w_3}=\frac{p-1}{2}p^{\frac{K-1}{2}}\big(p^{\frac{K-1}{2}}-1\big).
\end{align*} 

{ Case 2.} If $\frac{m_2}{v}$ is odd and $K$ is even, the length of $\mathcal{C}_{D_0}$ is $N_0=p^{K-1}+{L^{K}}(p-1)p^{\frac{K-2}{2}}-1$. It follows from $(\ref{N2})$ that the nonzero weights of $\mathcal{C}_{D_0}$ are 
\begin{align*}
w_1=(p-1)p^{\frac{K-2}{2}}\big(p^{\frac{K-2}{2}}+{L^{K}}\big),\qquad w_2=(p-1)p^{K-2}.
\end{align*}Then, by the first two Pless power moments, one has 
\begin{align*}
A_{w_1}=(p-1)p^{\frac{K-2}{2}}\big(p^{\frac{K}{2}}-{L^{K}}\big),\qquad A_{w_2}=p^{K-1}+{L^{K}}(p-1)p^{\frac{K-2}{2}}-1.
\end{align*}

{ Case 3.} If $\frac{m_2}{v}\equiv 2~(\mathrm{mod}~4)$ and $m_1$ is even, then the length of $\mathcal{C}_{D_0}$ is $N_0=p^{K-1}+{L^{m_1}}(p-1)p^{\frac{K-2}{2}}-1$, the nonzero weights of $\mathcal{C}_{D_0}$ are 
\begin{align*}
w_1=(p-1)p^{\frac{K-2}{2}}\big(p^{\frac{K-2}{2}}+{L^{m_1}}\big),\qquad w_2=(p-1)p^{K-2},
\end{align*}and by the first two Pless power moments, one has
\begin{align*}
A_{w_1}=(p-1)p^{\frac{K-2}{2}}\big(p^{\frac{K}{2}}-{L^{m_1}}\big),\qquad A_{w_2}=p^{K-1}+{L^{m_1}}(p-1)p^{\frac{K-2}{2}}-1.
\end{align*}

By the cases $1$-$3$, we can obtain Tables $1$-$3$, correspondingly, which combines Lemma \ref{l32} leads to the complete weight enumerator of $\mathcal{C}_{D_0}$ directly.

	So far, we complete the proofs for Theorems $\ref{t1}$-$\ref{t3}$.	$\hfill\Box$\\


{\bf The proofs for Theorems $\ref{t6}$-$\ref{t9}$}.

Using Lemmas $\ref{l31}$-$\ref{l35}$, in the similar proofs as those of Theorems $\ref{t1}$-$\ref{t3}$, correspondingly, one can
obtain Theorems $\ref{t6}$-$\ref{t9}$ immediately.	$\hfill\Box$

\subsection{The proofs for Theorems \ref{t4}-\ref{t5} and \ref{t10}-\ref{t11}}
In this subsection, Theorems \ref{t4}-\ref{t5} and \ref{t10}-\ref{t11} are obtained from Lemmas \ref{l12}, \ref{l31}-\ref{l32} and \ref{B}-\ref{A}. To this aim, we firstly prove Lemma $\ref{A}$.

{\bf Proof of Lemma \ref{A}}. 

If  $\frac{m_2}{v}~\equiv~0~(mod~4)$ and $m_1$ is even, by Lemmas \ref{l31}-\ref{l32} and \ref{B}, the length of $\mathcal{C}_{D_0}$ is $N_0-1=p^{K-1}+{L^{m_1}}(p-1)p^{\frac{K-2}{2}+v}-1$, the nonzero weights of $\mathcal{C}_{D_0}$ are 
\begin{align}\label{wc5}
\begin{aligned}
&w_1=(p-1)\big(p^{K-2}+{L^{m_1}}p^{\frac{K-2}{2}+v}\big),\\
&w_2=(p-1)p^{K-2},\\
&w_3=(p-1)\big(p^{K-2}+{L^{m_1}}(p-1)p^{\frac{K-4}{2}+v}\big),
\end{aligned}
\end{align} and
\begin{align}\label{awc5}	\begin{aligned}
&A_{w_1}=(p-1)p^{\frac{K-2}{2}-v}\big(p^{\frac{K}{2}-v}-{L^{m_1}}\big),\\
&A_{w_2}=p^{K-2v-1}+{L^{m_1}}(p-1)p^{\frac{K-2}{2}-v}-1,\\
& A_{w_3}=p^{K}\big(1-p^{-2v}\big).	\end{aligned}
\end{align} 
Then, by Lemma \ref{l32}, one has
\begin{align}\label{A0}
\#\bar{A}_0=A_{w_2}+1=p^{K-2v-1}+{L^{m_1}}(p-1)p^{\frac{K-2}{2}-v}.
\end{align}
Similarly, for $\lambda\in\mathbb{F}_{p}^{*}$, the length of $\mathcal{C}_{D_\lambda}$ is $N_0-1=p^{K-1}-{L^{m_1}}p^{\frac{K-2}{2}+v}$, the nonzero weights of $\mathcal{C}_{D_\lambda}$ are 
\begin{align}\label{wc10}
\begin{aligned}
&w_1=(p-1)p^{K-2},\\
&w_2=(p-1)\big(p^{K-2}-{L^{m_1}}p^{\frac{K-4}{2}+v}\big),\\
&w_3=(p-1)p^{K-2}-2{L^{m_1}}p^{\frac{K-2}{2}+v},
\end{aligned}
\end{align} and
\begin{align}\label{awc10}	\begin{aligned}
&A_{w_1}=\frac{p+1}{2}p^{K-2v-1}+\frac{p-1}{2}{L^{m_1}}p^{\frac{K-2}{2}-v}-1,\\
&A_{w_2}=p^{K}-p^{K-2v},\\
& A_{w_3}=\frac{p-1}{2}\big(p^{K-2v-1}-{L^{m_1}}p^{\frac{K-2}{2}-v}\big).	\end{aligned}
\end{align} 
Then, by Lemma \ref{l32}, one has
\begin{align}\label{Alambda}
\sum_{\substack{t\in\mathbb{F}_{p}^{*}\\\eta_{1}(t)=\eta_{1}(-\lambda)}}\#\bar{A}_t=A_{w_3}=\frac{p-1}{2}\big(p^{K-2v-1}-{L^{m_1}}p^{\frac{K-2}{2}-v}\big),
\end{align}
which leads to
\begin{align}\label{Alambda1}
\sum_{\substack{t\in\mathbb{F}_{p}^{*}\\\eta_{1}(t)=1}}\#\bar{A}_t=\sum_{\substack{t\in\mathbb{F}_{p}^{*}\\\eta_{1}(t)=-1}}\#\bar{A}_t=\frac{p-1}{2}\big(p^{K-2v-1}-{L^{m_1}}p^{\frac{K-2}{2}-v}\big).
\end{align}

For any given $\alpha\in\mathbb{F}_p$ with $\eta_{1}(\alpha)=-1$, we have 
\begin{align}\label{A+-}
\begin{aligned}		&\#\bar{A}_t\\=&\frac{1}{p}\sum_{a\in\mathbb{F}_{p^{m_1}}}\sum_{b\in B}\sum_{z\in\mathbb{F}_p}\zeta_p^{z\big(\mathrm{Tr}_{m_1}(\frac{a^2}{4})+\mathrm{Tr}_{m_2}(\gamma_b^{p^u+1})-t\big)}\\
=&p^{K-2v-1}+\frac{1}{p}\sum_{z\in\mathbb{F}_p^*}\zeta_p^{-tz}\sum_{a\in\mathbb{F}_{p^{m_1}}}\zeta_p^{\mathrm{Tr}_{m_1}(za^2)}\sum_{b\in B}\zeta_p^{z\mathrm{Tr}_{m_2}(\gamma_b^{p^u+1})}\\
=&p^{K-2v-1}+\frac{1}{p}G_{m_1}\sum_{z\in\mathbb{F}_p^*}\zeta_p^{-tz}\sum_{b\in B}\zeta_p^{z\mathrm{Tr}_{m_2}(\gamma_b^{p^u+1})}\\
=&p^{K-2v-1}+\frac{1}{p}G_{m_1}\bigg(\sum\limits_{\substack{z\in\mathbb{F}_p^{*}\\ \eta_{1}(z)=1}}\zeta_p^{-tz}\sum_{b\in B}\zeta_p^{z\mathrm{Tr}_{m_2}(\gamma_b^{p^u+1})}+\sum\limits_{\substack{z\in\mathbb{F}_p^{*}\\ \eta_{1}(z)=-1}}\zeta_p^{-tz}\sum_{b\in B}\zeta_p^{z\mathrm{Tr}_{m_2}(\gamma_b^{p^u+1})}\bigg)\\
=&p^{K-2v-1}+\frac{1}{p}G_{m_1}\bigg(\sum\limits_{\substack{z\in\mathbb{F}_p^{*}\\ \eta_{1}(z)=1}}\zeta_p^{-tz}\sum_{b\in B}\zeta_p^{\mathrm{Tr}_{m_2}(\gamma_b^{p^u+1})}+\sum\limits_{\substack{z\in\mathbb{F}_p^{*}\\ \eta_{1}(z)=-1}}\zeta_p^{-tz}\sum_{b\in B}\zeta_p^{\alpha\mathrm{Tr}_{m_2}(\gamma_b^{p^u+1})}\bigg)\\
=&p^{K-2v-1}+\frac{1}{p}G_{m_1}\bigg(\sum\limits_{z\in\mathbb{F}_p^{*}}\zeta_p^{-tz}\frac{(\eta_{1}(z)+1)}{2}A_{+}+\sum\limits_{z\in\mathbb{F}_p^{*}}\zeta_p^{-tz}\frac{(-\eta_{1}(z)+1)}{2} A_{-}\bigg)\\
=&p^{K-2v-1}+\frac{1}{2p}G_{m_1}\bigg(\big(A_{+}-A_{-}\big)\sum\limits_{z\in\mathbb{F}_p^{*}}\zeta_p^{-tz}\eta_{1}(z)+\big(A_{+}+A_{-}\big)\sum\limits_{z\in\mathbb{F}_p^{*}}\zeta_p^{-tz}\bigg),\\
\end{aligned}
\end{align}
where \begin{align*}
A_{+}=\sum_{b\in B}\zeta_p^{\mathrm{Tr}_{m_2}(\gamma_b^{p^u+1})},\qquad A_{-}=\sum_{b\in B}\zeta_p^{\alpha\mathrm{Tr}_{m_2}(\gamma_b^{p^u+1})}.
\end{align*}
Then, by $(\ref{A+-})$ and Lemma \ref{l21}, one has
\begin{align}\label{A01}
\#\bar{A}_0=p^{K-2v-1}-\frac{p-1}{2p}L^{m_1}p^{\frac{m_1}{2}}\big(A_{+}+A_{-}\big),
\end{align}
and for $t \in\mathbb{F}_{p}^*$,  
\begin{align}\label{A0lambda}
\#\bar{A}_t=p^{K-2v-1}+\frac{1}{2p}L^{m_1}p^{\frac{m_1}{2}}\big(\eta_{1}(-t)G_1(A_{+}-A_{-})-(A_{+}+A_{-})\big).
\end{align}		
It follows from $(\ref{A0})$, $(\ref{Alambda1})$ and $(\ref{A01})$-$(\ref{A0lambda})$ that
\begin{align}\label{A2}
A_{+}=A_{-}=-p^{\frac{m_2}{2}-v}.
\end{align}
Now by $(\ref{A01})$-$(\ref{A2})$, we can obtain $(\ref{m1even})$. 

Similarly, if $m_1$ is odd, by $(\ref{A2})$, we have
\begin{align}\label{A+-1}
\begin{aligned}		&\#\bar{A}_t\\
=&p^{K-2v-1}+\frac{1}{p}G_{m_1}\bigg(\sum\limits_{\substack{z\in\mathbb{F}_p^{*}\\ \eta_{1}(z)=1}}\zeta_p^{-tz}\eta_{1}(z)\sum_{b\in B}\zeta_p^{z\mathrm{Tr}_{m_2}(\gamma_b^{p^u+1})}+\sum\limits_{\substack{z\in\mathbb{F}_p^{*}\\ \eta_{1}(z)=-1}}\zeta_p^{-tz}\eta_{1}(z)\sum_{b\in B}\zeta_p^{z\mathrm{Tr}_{m_2}(\gamma_b^{p^u+1})}\bigg)\\
=&p^{K-2v-1}+\frac{1}{p}G_{m_1}\bigg(\sum\limits_{z\in\mathbb{F}_p^{*}}\zeta_p^{-tz}\frac{(\eta_{1}(z)+1)}{2}A_{+}-\sum\limits_{z\in\mathbb{F}_p^{*}}\zeta_p^{-tz}\frac{(-\eta_{1}(z)+1)}{2} A_{-}\bigg)\\
=&p^{K-2v-1}+\frac{1}{2p}G_{m_1}\bigg(\big(A_{+}+A_{-}\big)\sum\limits_{z\in\mathbb{F}_p^{*}}\zeta_p^{-tz}\eta_{1}(z)+\big(A_{+}-A_{-}\big)\sum\limits_{z\in\mathbb{F}_p^{*}}\zeta_p^{-tz}\bigg)\\
=&\begin{cases}
p^{K-2v-1},\qquad&t=0;\\
p^{K-2v-1}-\eta_{1}(-t){L^{m_1+1}}p^{\frac{K-1}{2}-v},\qquad&\text{otherwise}.
\end{cases}
\end{aligned}
\end{align}

So far, we complete the proof of Lemma \ref{A}.$\hfill\Box$\\


{\bf The proofs for Theorems \ref{t4}-\ref{t5} and \ref{t10}-\ref{t11}}. 

If $\frac{m_2}{v}\equiv 0~(\mathrm{mod}~4)$ and $m_1$ is even,  Tables $4$ and $10$ are given from $(\ref{wc5})$-$(\ref{awc5})$ and  $(\ref{wc10})$-$(\ref{awc10})$, respectively, and then the complete weight enumerators of $\mathcal{C}_{D_0}$ and $\mathcal{C}_{D_\lambda}$ are obtained from lemmas $\ref{l32}$ and $\ref{A}$ directly. So far, we complete the proofs for Theorems $\ref{t4}$ and $\ref{t10}$. 

By Lemmas $\ref{l31}$-$\ref{l32}$ and $\ref{B}$-$\ref{A}$, in the similar proofs as those of Theorems $\ref{t4}$ and $\ref{t10}$, we can
obtain Theorems  $\ref{t5}$ and $\ref{t11}$ immediately.	$\hfill\Box$\\
\section{Examples}

In this section, we give some examples for the main results.
\begin{example}
	For $p=3$, by using Magma, we obtain $\mathcal{C}_{D_0}$ for some special cases in Table $12$, which are accordant with Theorems $\ref{t1}$-$\ref{t5}$.
	\begin{center} Table $12$~~~Some $\mathcal{C}_{D_0}$ for special cases
		
		\scalebox{1.1}{	\begin{tabular}{|p{0.6cm}<{\centering}|p{0.6cm}<{\centering}|p{0.6cm}<{\centering}|p{0.6cm}<{\centering}|p{1.5cm}<{\centering}|p{2.5cm}<{\centering}| p{5cm}<{\centering}|}
				\hline  $m_1$                  & $m_2$  &   $u$  & $\frac{m_2}{v}$  &$m_1+m_2$  & parameter       &   weight enumerator                    \\ 
				\hline  $3$                    & $2$    &   $2$  &  $1$ &$5$  &\small$[80,5,48]$      &  \small$1+90z^{48}+80z^{54}+72{z^{60}}$                   \\ 		
				\hline  $2$                    & $2$    &   $2$  &  $1$ &$4$  &\small$[32,4,18]$      &  \small$1+32z^{18}+48z^{24}$                   \\ 	
				\hline  $2$                    & $2$    &   $1$  &  $2$ &$4$   &\small$[20,4,12]$      &  \small$1+60z^{12}+20z^{18}$                   \\
				\hline  $2$                    & $4$    &   $2$  &  $2$ &$4$   &\small$[224,6,144]$      &  \small$1+504z^{144}+224z^{162}$                   \\ 		
				\hline  $2$                    & $4$    &   $3$  &  $4$ &$6$  &\small$[188,6,108]$      &  \small$1+60z^{108}+648z^{126}+20z^{162}$        \\ 
				\hline  $3$                    & $4$    &   $1$  &  $4$ &$7$   &\small$[728,7,432]$      &  \small$1+90z^{432}+2024z^{486}+72z^{540}$                   \\ 
				\hline
		\end{tabular} }
	\end{center} 
\end{example}

According to the Griesmer bound \cite{JH1960}, the code $[20,4,12]$ is optimal. 
\begin{example}
	For $p=3$, by Table $12$ and Theorems $\ref{t1}$-$\ref{t5}$, we obtain Furthermore, $\mathcal{C}_{\tilde{D}_0}$ for some special cases in Table $12^{\circ}$.
	\begin{center} Table $12^{\circ}$ ~~~Some Furthermore, $\mathcal{C}_{\tilde{D}_0}$ for special cases
		
		\scalebox{1.1}{	\begin{tabular}{|p{0.6cm}<{\centering}|p{0.6cm}<{\centering}|p{0.6cm}<{\centering}|p{0.6cm}<{\centering}|p{1.4cm}<{\centering}|p{2cm}<{\centering}| p{5cm}<{\centering}|}
				\hline  $m_1$                  & $m_2$  &   $u$  & $\frac{m_2}{v}$  &$m_1+m_2$  & parameter       &   weight enumerator                    \\ 
				\hline  $3$                    & $2$    &   $2$  &  $1$ &$5$  &\small$[40,5,24]$      &  \small$1+90z^{24}+80z^{27}+72{z^{30}}$                   \\ 		
				\hline  $2$                    & $2$    &   $2$  &  $1$ &$4$  &\small$[16,4,9]$      &  \small$1+32z^{9}+48z^{12}$                   \\ 	
				\hline  $2$                    & $2$    &   $1$  &  $2$ &$4$   &\small$[10,4,6]$      &  \small$1+60z^{6}+20z^{9}$                   \\
				\hline  $2$                    & $4$    &   $2$  &  $6$ &$4$   &\small$[112,6,72]$      &  \small$1+504z^{72}+224z^{81}$                   \\ 		
				\hline  $2$                    & $4$    &   $3$  &  $4$ &$6$  &\small$[94,6,54]$      &  \small$1+60z^{54}+648z^{63}+20z^{81}$        \\ 
				\hline  $3$                    & $4$    &   $1$  &  $4$ &$7$   &\small$[364,7,216]$      &  \small$1+90z^{216}+2024z^{243}+72z^{270}$                   \\ 
				\hline
		\end{tabular} }
	\end{center} 
\end{example}

According to the Griesmer bound \cite{JH1960}, the codes $[16,4,9]$, $[10,4,6]$ and $[112,6,72]$ are all optimal. 

\begin{example}
	For $p=3$, by using Magma, we obtain $\mathcal{C}_{D_\lambda}$ for some special cases in Table $13$, which are accordant with Theorems $\ref{t6}$-$\ref{t11}$.
	\begin{center} Table $13$~~~ Some $\mathcal{C}_{D_\lambda}$ for special cases
		
		\scalebox{1.1}{	\begin{tabular}{|p{0.5cm}<{\centering}|p{0.5cm}<{\centering}|p{0.5cm}<{\centering}|p{0.5cm}<{\centering}|p{0.5cm}<{\centering}|p{1.4cm}<{\centering}|p{2cm}<{\centering}| p{5cm}<{\centering}|}
				\hline  $\lambda$      & $m_1$                  & $m_2$  &   $u$  & $\frac{m_2}{v}$  &$m_1+m_2$  & parameter       &   weight enumerator                    \\ 
				\hline  $-1$&$3$                    & $2$    &   $2$  &  $1$ &$5$  &\small$[90,5,54]$      &  \small$1+80z^{54}+72z^{60}+90{z^{66}}$                   \\ 		
				\hline  $1$&$2$                    & $2$    &   $2$  &  $1$ &$4$  &\small$[24,4,12]$      &  \small$1+24z^{12}+56z^{18}$                   \\ 	
				\hline  $-1$&$2$                    & $2$    &   $1$  &  $2$ &$4$   &\small$[30,4,18]$      &  \small$1+50z^{18}+30z^{24}$                   \\
				\hline  $1$&$2$                    & $4$    &   $2$  &  $2$ &$4$   &\small$[252,6,162]$      &  \small$1+476z^{162}+252z^{180}$                   \\ 		
				\hline  $-1$&$2$                    & $4$    &   $1$  &  $4$ &$6$  &\small$[270,6,162]$      &  \small$1+50z^{162}+648z^{180}+30z^{216}$        \\ 
				\hline  $-1$&$3$                    & $4$    &   $1$  &  $4$ &$7$   &\small$[648,7,378]$      &  \small$1+72z^{378}+2034z^{432}+80z^{486}$                   \\ 
				\hline
		\end{tabular} }
	\end{center} 
\end{example}

According to the Griesmer bound \cite{JH1960}, the code $[30,4,18]$ is almost optimal. 
\begin{example}
	For $p=3$, by Table $13$ and Theorems $\ref{t6}$-$\ref{t11}$, we obtain $\mathcal{C}_{\tilde{D}_\lambda}$ for some special cases in Table $13^{\circ}$.
	\begin{center} Table $13^{\circ}$~~~ Some $\mathcal{C}_{\tilde{D}_\lambda}$ for special cases
		
		\scalebox{1.1}{	\begin{tabular}{|p{0.5cm}<{\centering}|p{0.5cm}<{\centering}|p{0.5cm}<{\centering}|p{0.5cm}<{\centering}|p{0.5cm}<{\centering}|p{1.4cm}<{\centering}|p{2cm}<{\centering}| p{5cm}<{\centering}|}
				\hline  $\lambda$      & $m_1$                  & $m_2$  &   $u$  & $\frac{m_2}{v}$  &$m_1+m_2$  & parameter       &   weight enumerator                    \\ 
				\hline  $-1$&$3$                    & $2$    &   $2$  &  $1$ &$5$  &\small$[45,5,27]$      &  \small$1+80z^{27}+72z^{30}+90{z^{33}}$                   \\ 		
				\hline  $1$&$2$                    & $2$    &   $2$  &  $1$ &$4$  &\small$[12,4,6]$      &  \small$1+24z^{6}+56z^{9}$                   \\ 	
				\hline  $-1$&$2$                    & $2$    &   $1$  &  $2$ &$4$   &\small$[15,4,9]$      &  \small$1+50z^{9}+30z^{12}$                   \\
				\hline  $1$&$2$                    & $4$    &   $2$  &  $2$ &$4$   &\small$[126,6,81]$      &  \small$1+476z^{81}+252z^{90}$                   \\ 		
				\hline  $-1$&$2$                    & $4$    &   $1$  &  $4$ &$6$  &\small$[135,6,81]$      &  \small$1+50z^{81}+648z^{90}+30z^{108}$        \\ 
				\hline  $-1$&$3$                    & $4$    &   $1$  &  $4$ &$7$   &\small$[324,7,189]$      &  \small$1+72z^{189}+2034z^{216}+80z^{243}$                   \\ 
				\hline
		\end{tabular} }
	\end{center} 
\end{example}

According to the Griesmer bound \cite{JH1960}, the codes $[45,5,27]$ and $[12,4,6]$ are both almost optimal, $[15,4,9]$ and $[126,6,81]$ are both optimal. 
\section{Conclusion}
Note that codes in \cite{CL2016, GJ2019} are always even dimension. In this paper, we construct several classes of two-weight and three-weight linear codes with any dimension over the finite field $\mathbb{F}_p$ ($p$ is an odd prime) by extending the construction in \cite{CL2016, GJ2019},  and we determine their complete weight enumerators by using Weil sums. Furthermore, according to the Griesmer bound, some examples of these codes are optimal or almost optimal, respectively.
	 			

\begin{thebibliography}{10}
	\bibitem{AC1984} A. Calderbank, J. Goethals, Three-weight codes and association schemes, Philips J. Res. 39(4-5) (1984) 143-152.
	\bibitem{RC1986} R. Calderbank, W. Kantor, The geometry of two-weight codes, Bull. Lond. Math. Soc. 18(2) (1986) 97-122.
	\bibitem{CC2005} C. Carlet, C. Ding, J. Yuan, Linear codes from perfect nonlinear mappings and their secret sharing schemes, IEEE Trans. Inf. Theory. 51(6) (2005) 2089-2102.
	\bibitem{CR} R. Coulter, Explicit evaluations of some Weil sums, Acta Arith. 83 (1998) 241-251.
    \bibitem{CR1} R. Coulter, Further evaluations of Weil sums, Acta Arith. 86 (1998) 217-226.
	\bibitem{CD2005}C. Ding, X. Wang, A coding theory construction of new systematic authentication codes, Theor. Comput. Sci. 330(1) (2005) 81-99.
	\bibitem{KD2014}K. Ding, C. Ding, Binary linear codes with three weights, IEEE Commun. Lett. 18(11) (2014) 1879-1882.
	\bibitem{KD2015}K. Ding, C. Ding, A class of two-weight and three-weight codes and their applications in secret sharing, IEEE Trans. Inf. Theory. 61(11) (2015) 5835-5842.
	\bibitem{JH1960}J. Griesmer, A bound for error-correcting codes, IBM J. Res. Dev. 4(5) (1960) 532-542.
		
	\bibitem{WV} W. Huffman, V. Pless, Fundamentals of Error-Correcting Codes, Cambridge University Press. 2010.
	\bibitem{ZH2015}Z. Heng, Q. Yue, A class of binary linear codes with at most three weights, IEEE Commun. Lett. 19(9) (2015) 1488-1491.
	\bibitem{ZH2016}Z. Heng, Q. Yue, Two classes of two-weight linear codes, Finite Fields Appl. 38 (2016) 72-92.
	\bibitem{ZH20161}Z. Heng, Q. Yue, A construction of $q$-ary linear codes with two weights, Finite Fields Appl. 48 (2017) 20-42.
	\bibitem{ZH20162}Z. Heng, Q. Yue, C. Li, Three classes of linear codes with two or three weights, Discrete Math. 339(11) (2016) 2832-2847.
	\bibitem{GJ2019}G. Jian, Z. Lin, R. Feng, Two-weight and three-weight linear codes based on Weil sums, Finite Fields Appl. 57 (2019) 92-107.
	\bibitem{CS2019} C. Li,  S. Bae, S. Yang,  Some two-weight and three-weight linear codes, Advances in Mathematics of Communications. 13(1) (2019) 195-211.
	\bibitem{CL2016}C. Li, Q. Yue, F. Fu, A construction of several classes of two-weight and three-weight linear codes, Appl. Algebra Eng. Commun. Comput. (2016) 1-20.
	\bibitem{RL97} R. Lidl,  H. Niederreiter, Cohn F.M.. Finite Fields. Cambridge University Press, Cambridge. (1997).
	\bibitem{GL2018}G. Luo, X. Cao, S. Xu, J. Mi, Binary linear codes with two or three weights from niho exponents, Cryptogr. Commun. 10(2) (2018) 301-318	.
	\bibitem{CT2016}C. Tang, N. Li, Y. Qi, Z. Zhou, T. Helleseth, Linear codes with two or three weights from weakly regular bent functions, IEEE Trans. Inf. Theory.	 62(3) (2016) 1166-1176.
	\bibitem{KT2007} K. Torleiv, Codes for Error Detection, vol. 2, World Scientific. (2007).
	\bibitem{JY2006} J. Yuan, C. Ding, Secret sharing schemes from three classes of linear codes, IEEE Trans. Inf. Theory. 52(1) (2006) 206-212.
	\bibitem{SY2015} S. Yang, Z. Yao, Complete weight enumerators of a family of three-weight linear codes, Des. Codes Cryptogr. (2017) 663-674.        
	\bibitem{ZZ2015}Z. Zhou, N. Li, C. Fan, T. Helleseth, Linear codes with two or three weights from quadratic bent functions, Des. Codes Cryptogr. (2015) 1-13.
\end{thebibliography}
       \end{document}